\documentclass[useAMS, usenatbib, usegraphicx]{mn2e}

 \usepackage{url}
 \usepackage{lscape}
 \usepackage{amssymb}

\title[Optical spectra of 5 new SMC BeXRBs]{Optical spectra of 5 new Be/X-ray Binaries in the Small Magellanic Cloud and the link of the supergiant B$[$e$]$ star LHA 115-S 18 with an X-ray source}

\author[G. Maravelias et al.]
{
 \parbox{\textwidth}{
  G. Maravelias,$^{1}$\thanks{E-mail:\texttt{gmaravel@physics.uoc.gr}}
  A. Zezas,$^{1,2,3}$
  V. Antoniou,$^{3,4}$
  D. Hatzidimitriou$^{5}$
 }
 \vspace{0.4cm}\\
 \parbox{\textwidth}{
  $^{1}$University of Crete, Physics Department \& Institute of Theoretical \& Computational Physics, GR-710 03 Heraklion, Crete, Greece\\
  $^{2}$Foundation for Research and Technology-Hellas, Institute of Electronic Structure \& Laser, GR-711 10 Heraklion, Crete, Greece\\
  $^{3}$Harvard-Smithsonian Center for Astrophysics, 60 Garden Street, Cambridge, MA 02138, USA\\
  $^{4}$Iowa State University, Department of Physics \& Astronomy, Ames, IA 50011, USA \\
  $^{5}$University of Athens, Department of Physics, Section of Astrophysics, Astronomy, and Mechanics, GR-157 84 Zografou, Athens, Greece
 }
}

\begin{document}

\date{Accepted 2013 November 27. Received 2013 November 25; in original form 2013 February 01}

\pagerange{\pageref{firstpage}--\pageref{lastpage}} \pubyear{2012}

\maketitle

\label{firstpage}

\begin{abstract}

The Small Magellanic Cloud (SMC) is well known to harbor a large number of High-Mass X-ray Binaries (HMXBs). The identification of their optical counterparts provides information on the nature of the donor stars and can help to constrain the parameters of these systems and their evolution. We obtained optical spectra for a number of HMXBs identified in previous \textit{Chandra} and \textit{XMM-Newton} surveys of the SMC using the AAOmega/2dF fiber-fed spectrograph at the Anglo-Australian Telescope. We find 5 new Be/X-ray binaries (BeXRBs; including a tentative one), by identifying the spectral type of their optical counterparts, and we confirm the spectral classification of an additional 15 known BeXRBs. We compared the spectral types, orbital periods, and eccentricities of the BeXRB populations in the SMC and the Milky Way and we find marginal evidence for difference between the spectral type distributions, but no statistically significant differences for the orbital periods and the eccentricities. Moreover, our search revealed that the well known supergiant B[e] star LHA 115-S 18 (or AzV 154) is associated with the weak X-ray source CXOU J005409.57-724143.5. We provide evidence that the supergiant star LHA 115-S 18 is the optical counterpart of the X-ray source, and we discuss different possibilities of the origin of its low X-ray luminosity (Lx $\sim4\times10^{33}$ erg s$^{-1}$).
\end{abstract}
 
\begin{keywords}
galaxies: individual (Galaxy) -- Magellanic Clouds -- stars: emission-line, Be -- X-rays: binaries -- star: individual: LHA 115-S 18
\end{keywords}

\section{Introduction}
\label{s-introduction}

High-Mass X-ray Binaries (HMXBs) are stellar systems consisting of a massive, early-type star (of O or B spectral type) and a compact object (neutron star or black hole). The material lost by the companion star (the donor) either through strong stellar winds or a circumstellar disk is accreted onto the compact object resulting in the formation of supergiant X-ray Binaries (sgXRBs) and Be/X-ray Binaries (BeXRBs) respectively. Predominantly the compact objects in these systems are pulsars with spin periods in the 1-1000 s range (e.g. \citealt{Knigge11}). Depending on the available material and the geometry of the orbit of the systems, their X-ray emission can be either persistent or variable in timescales of days up to several months. Their typical luminosity ranges between $\sim10^{34}$ (for low-activity systems) up to $10^{38}$ erg s$^{-1}$ (for outbursting systems). 

In the BeXRBs, the most numerous subclass of HMXBs \citep{Liu05,Liu06}, the donor is a non-supergiant B star (luminosity class III-V) whose spectrum shows or, has at some time in the past shown, Balmer lines in emission (the so-called "Be phenomenon"; e.g. \citealt{Porter03}). This emission is produced by an equatorial disk of ionized material that has been expelled from the star due to its high (close to the critical limit) rotational velocity. Subsequently, part of this material is accreted on the compact object. Most BeXRBs are transient systems (e.g. \citealt{Reig99}) which can produce outbursts with luminosities in the range of $10^{36}-10^{37}$ erg s$^{-1}$ (type I outbursts, which occur at periastron and last a few days) or even stronger with luminosities $\gtrsim10^{37}$ erg s$^{-1}$ (type II outbursts, which are more rare and occur at irregular intervals). 
However, there are also persistent sources which display lower luminosity levels ($\sim10^{34}-10^{35}$ erg s$^{-1}$; for a review see \citealt{Reig11}).

In the case of sgXRBs (\citealt{Charles06,Liu06}, and references therein), the donor is a supergiant O or B-type star (luminosity class I-II). Depending on the mass-transfer mechanism these systems are divided in Roche-lobe overflow (RLOF; e.g. \citealt{Lamers76}) and wind-fed (e.g. \citealt{Lutovinov13}). For systems in the first subclass, the steady mass-transfer rate through the Roche-lobe is high enough to lead to the formation of an accretion disk around the compact object, resulting in persistent systems with luminosities up to $\sim10^{38}$ erg s$^{-1}$.           
In wind-fed systems, the donor star loses mass through a strong radial stellar wind (with mass-loss rates between $10^{-8}-10^{-6}$ M$_\odot$ yr$^{-1}$). As the compact object lies in a close orbit around the donor, it becomes a persistent X-ray source with much lower luminosity (in the range $\sim10^{35}-10^{36}$ erg s$^{-1}$). These systems are referred as "classical". The advent of \textit{INTEGRAL} has unveiled new populations of wind-fed systems: the supergiant Fast X-ray Transients (SFXTs; \citealt{Negueruela06}) and the heavily obscured sgXRBs \citep{Walter06}. The SFXTs display flaring activity which lasts from few minutes to several hours with a luminosity increase from $\sim10^{33}-10^{34}$ erg s$^{-1}$ to $\sim10^{36}-10^{37}$ erg s$^{-1}$. On the other hand, the heavily obscured sgXRBs are actually similar to the "classical" wind-fed sgXRBs but the compact object is deeply embedded in a dense absorbing environment. Their ${\rm H_{I}}$ column density can be as high as $N_H\sim10^{24}$ cm$^{-2}$ thus suppressing significantly their observed X-ray luminosities. For comparison the measured absorbing column density for the "classical" systems is of the order of $\sim10^{21}-10^{22}$ cm$^{-2}$ (for a review see \citealt{Kaper04}; \citealt{Chaty11}).

The Small Magellanic Cloud (SMC) is an excellent laboratory to study the HMXBs, since it is nearby (D = 60 kpc; \citealt{Hilditch05}) and well-covered by the  \textit{Chandra} and \textit{XMM-Newton} X-ray observatories that can detect sources down to ${\rm L_X}\sim 10^{33}$ erg s$^{-1}$ (i.e. reaching luminosities of non-outbursting sources). Moreover, it does not suffer from large extinction and distance uncertainties that often hamper studies of HMXBs in the Milky Way. It also has a relatively uniform metallicity among the young populations, and a well-determined star-formation history \citep{Harris04}. Most importantly it is host to a large number ($\sim90$) of HMXBs (\citealt{Haberl04,Coe05,Antoniou09}, 2009b, 2010).
Out of these systems only one is a sgXRB, source SMC X-1 \citep{Webster72}, which is the only persistent accreting X-ray pulsar (P$_{spin}\sim0.71$ s; \citealt{Lucke76}) in the SMC fed through RLOF. This system has a B0 supergiant companion \citep{Webster72} with an orbital period of 3.89 days \citep{Tuohy75} and an X-ray luminosity of $\sim9.5\times10^{37}$ erg s$^{-1}$ in the 0.2-12.0 keV energy band (e.g. XMMSL1; \citealt{Saxton08}). In contrast, in the Milky Way the number of confirmed or suspected supergiant systems is much higher ($\sim32$\% of the total number of HMXBs; \citealt{Liu06,Chaty11}).

Although the number of known HMXBs in the SMC has increased dramatically in the last decade, only recently we started having a picture of the spectral classification of their donor stars \citep[e.g.][]{McBride08,Antoniou09}. This is important since it can yield valuable information on the evolution of massive binary stellar systems. 
Following our previous work \citep{Antoniou09}, we used the multiple-object mode of the AAOmega spectrograph, a fiber-fed optical spectrograph on the 3.9m Anglo-Australian Telescope (AAT), to obtain optical spectra of confirmed and candidate HMXBs, in order to identify BeXRBs and determine their spectral types.
 
In this paper we present the results of this spectroscopic campaign. Its structure is the following: In Section \ref{s-sample} we describe the sample of sources and in Section \ref{s-obs+dataanalysis} we discuss the observations and the data reduction. In Section \ref{s-select_bexrbs} we present the selection criteria of candidate BeXRBs. In Section \ref{s-spectral_class} the spectral classification of the BeXRBs and the comparison of their spectral types with previous results are discussed. In Section \ref{s-discussion} we discuss the properties of the overall population of BeXRBs in the SMC and we compare them with the BeXRB population in the Milky Way. We also discuss the nature of the X-ray source CXOU J005409.57-724143.5 that is associated with the supergiant B[e] star LHA 115-S 18 (hereafter S 18, \citealt{Henize56}; also known as AzV 154; \citealt{Azzopardi75}). A summary of the main results of this study is given in Section \ref{s-conclusions}.  

\section{Sample}
\label{s-sample}

The sample used in this work is derived from studies of X-ray sources detected with the \textit{Chandra} and \textit{XMM-Newton} X-ray observatories. As our basic sample we use the catalog of HMXB candidates detected in the \textit{Chandra} shallow survey of the SMC \citep{Antoniou09bb}, which were identified based on the location of their optical counterparts in the  (\textit{V, B-V}) color-magnitude diagram (CMD). The chance-coincidence probability for a \textit{Chandra} X-ray source to be associated with an OB star is estimated to be $\sim$20\% \citep{Antoniou09bb}. This approach allowed us to identify candidate HMXBs even when we could not detect X-ray pulsations in the X-ray data, a tell-tale signature of BeXRB pulsars. It also allowed us to identify objects of lower X-ray luminosities than it would be impossible based on the detection of X-ray pulsations.  

In this work we use the sample of \citet{Antoniou09bb}, which includes the most likely optical counterpart of 158 \textit{Chandra} sources with X-ray luminosities as low as ${\rm L_X}\sim4\times10^{33}$ erg s$^{-1}$, of uncertain or unknown spectral types. Moreover, this sample is supplemented by 211 additional sources detected in various \textit{XMM-Newton} observations of the SMC reaching ${\rm L_X}\sim3.5\times10^{33}$ erg s$^{-1}$ (\citealt{Haberl04,Antoniou10}), which also have uncertain or unpublished spectral types. We were able to obtain spectra for 133 \textit{Chandra} and 151 \textit{XMM-Newton} sources in total.

\section{Observations and data analysis}
\label{s-obs+dataanalysis}

{
\setlength{\tabcolsep}{2pt}
\begin{table*}
 \caption{Summary of the service-time observing runs with the AAOmega spectrograph.}
 \begin{tabular}{lccccccccc}
 \hline\hline
  Field ID & \multicolumn{2}{c}{Field center} & Observation & Exposure & Grating & $\lambda$ Range & Dispersion & Resolution & Allocated objects$^*$\\
     & \multicolumn{2}{c}{R.A. (J2000) Dec.} & date & & & &  \\
     & (h m s) & ($^{o}$ ' ") & & (s) & & ({\AA}) & ({\AA}/pix) & {\AA} & \\
 \hline
  26jul08\_north & 01 00 20 & -72 25 25 & 26/7/2008 & 3x1800 & 580V & 3733.0-5857.3 & 1.03 & 3.21 & 74(C),68(X),46(s) \\
                 &          &           &           &        & 385R & 5579.6-8808.9 & 1.57 & 5.70 \\                 
  26jul08\_south & 00 43 10 & -73 08 49 & 26/7/2008 & 4x1800 & 580V & 3733.0-5857.3 & 1.03 & 3.21 & 59(C),78(X),25(s) \\
                 &          &           &           &        & 385R & 5579.6-8808.9 & 1.57 & 5.70 \\                 
  19sep08\_south & 01 00 20 & -72 25 25 & 19/9/2008 & 8x1200 & 580V & 3682.6-5807.3 & 1.03 & 3.21 & 59(C),77(X),25(s) \\
                 &          &           &           &        & 1000R & 5907.5-7080.0 & 0.57 & 1.94 \\                 
 \hline
 \end{tabular}

 \begin{flushleft}
  $^*$ The allocated objects are \textit{Chandra} sources (labeled as 'C'), \textit{XMM-Newton} sources (labeled as 'X'), dedicated sky fibers (labeled as 's').\\
  Unique observed objects in north field: 74 \textit{Chandra} and 68 \textit{XMM-Newton} sources.\\ 
  Unique observed objects in south fields: 59 \textit{Chandra} and 72 \textit{XMM-Newton} sources. Additionaly, there were 6 and 5 more \textit{XMM-Newton} sources observed in 26jul08\_south and 19sep08\_south fields, respectively. 
 \end{flushleft}
\label{t-obslog}
\end{table*}
}

\subsection{AAOmega spectroscopy}
\label{s-aaomega_obs}
Although optical photometry for these sources has identified them as candidate HMXBs, and it is a powerful tool for identifying large samples of such objects, only optical spectroscopy can unambiguously confirm this classification, and provide additional information on the nature of these systems. 

The optical spectra for this study were acquired during two nights of service time (on July 26 and September 19, 2008), using the multi-object mode of the AAOmega spectrograph \citep{Sharp06}, a double-arm fiber-fed optical spectrograph (up to 400 fibers) on the 3.9m Anglo-Australian Telescope (AAT) fed by the 2 Degree Field (2dF) robotic fiber positioner. 
A summary of the observing runs is presented in Table \ref{t-obslog}. In addition, flat-field and arc (FeAr+CuAr+CuHe+CuNe) calibration exposures were taken each night for each setup.

The data reduction was performed with the 2dfdr\footnote{\url{http://www.aao.gov.au/2df/aaomega/aaomega_2dfdr.html}} v4 package with default values. However, we did not perform the sky subtraction built in 2dfdr as we followed a different approach than the standard procedure. The steps taken during the 2dfdr process included: (i) bias subtraction and flat-fielding; (ii) wavelength calibration; (iii) combination of individual exposures for the same field. 
The extraction of the individual spectra was performed with the \textit{extract} command of the FIGARO v5.6-6 package of STARLINK \citep{Shortridge04}. 

After the extraction of all spectra from the AAOmega data, we performed the sky subtraction and initial characterization of the spectra. Flux calibration was not attempted since the wavelength-dependent throughput of each fiber is different and, ideally, a flux standard should be observed through each fiber. Nevertheless, our analysis is not affected because our classification criteria are based mainly on the presence or absence of spectral lines and not their absolute intensity. In addition, we can use the relative intensity of nearby lines, as the majority of the lines used in the classification are between 3900 {\AA} and 4700 {\AA} (with the exception of the H$\alpha$ line), where the fiber response is fairly flat.

We considered for further analysis objects with S/N ratio above 20 (i.e. 400 counts) in the blue (4100 {\AA} - 4300 {\AA}) as well as the red (6410 {\AA} - 6450 {\AA}) parts of the spectrum.
For the sky spectra, since the final sky spectrum resulted from the combination of several spectra, we set a limiting S/N ratio of 15 (i.e. 225 counts) in each of the two bands. After this selection, a total of 25 sky and 130 source spectra were kept from the September 19 observing run, while from the July 26 run we kept 5 sky and 53 spectra for further analysis.

The field of each sky fiber was examined visually (using the images from the OGLE-II project\footnote{\url{http://ogledb.astrouw.edu.pl/~ogle/photdb/}}; \citealt{Udalski98}) in order to ensure that the spectra were not contaminated by any nearby source. After this process, 10 spectra were selected for further analysis from the south field (9 observed on September 19 and 1 on July 26) and 2 from the north field (observed on July 26). These sky spectra were combined into a median sky spectrum for each observation date and field. 

All object and sky spectra were corrected for small residual wavelength offsets by measuring the positions of strong sky emission lines at 5577.3 {\AA} and 6300.3 {\AA} for the blue and the red band respectively. In order to account for throughput variations between the object and sky fibers, the fluxes for these sky emission lines were measured for each stellar spectrum and the corresponding sky spectrum was scaled in order to match the measured intensity of the lines. Then the rescaled sky spectrum was subtracted from the corresponding object spectrum. 

Although this method is sufficient for the subtraction of sky emission from each spectrum, it is not sufficient for the removal of the contaminating interstellar emission in the stellar spectra. The selected fields in the SMC show strong and spatially variable diffuse emission from ${\rm H_{II}}$ regions and supernova remnants. Among the interstellar emission lines, H$\alpha$ is the strongest one but, at the same time, it is also a critical feature for the classification of BeXRBs. Ideally, sky subtraction would be performed with sky fibers placed within few arcsec from each source, in order to correct for the local interstellar contamination. However, due to hardware limitations we cannot place two fibers closer than 30\arcsec\footnote{Although the minimum distance is 30 arcsec, a typical distance is closer to 30-40 arcsec (\url{http://www.aao.gov.au/AAO/2df/aaomega/aaomega_faq.html#fibsep}).}, leaving us with the only option of measuring an average diffuse emission spectrum for each field. 

By measuring a mean sky spectrum, we can remove a large part of the interstellar emission background but there may still be some residual contamination. This is indicated by the presence of typical interstellar emission lines (such as [OIII] $\lambda$5007 and [SII] $\lambda\lambda$6716,6731) in the sky subtracted spectra.

\subsection{Optical and Infrared data for star S 18}
\label{s-optir_data}

Optical photometry for star S 18 (\citealt{Henize56}, or AzV 154 after \citealt{Azzopardi75}) has been derived from the Optical Gravitational Lensing Experiment (OGLE) online database (see footnote 2; \citealt{Udalski97,Szymanski05}). 
The retrieved data were obtained in the Bessell $I$-band. There are 327 observations between June 1997 and November 2000, with typical photometric errors of 0.003 mag.
Although photometry for this star also exists in the MAssive Compact Halo Object (MACHO) database \citep{Alcock97,Alcock99}, these data show unphysically large scale scatter of up to $\sim$1 mag compared to the OGLE data. We attribute this to confusion with a star of similar brightness located $\sim4\arcsec$ from S 18  (for comparison the median seeing of the MACHO survey is $\sim3\arcsec$; \citealt{Alcock97}).

We used the catalog compiled by \citet{Bonanos10} to retrieve the infrared photometric properties of star S 18 (discussed in Section \ref{s-sgcase_optirprop}): $J$=12.349$\pm$0.033 mag, $H$=11.931$\pm$0.038 mag, $K_{s}$=11.109$\pm$0.026 mag, [3.6$\mu$m]=9.177$\pm$0.042 mag, [8$\mu$m]=6.966$\pm$0.022 mag, [24$\mu$m]=4.786$\pm$0.007 mag. 

\subsection{X-ray data for source CXOU J005409.57-724143.5}
\label{s-xray_data}

In the $\sim$9.4 ks long \textit{Chandra} observation obtained in July 04, 2002, we detected this X-ray source with an absorption corrected X-ray luminosity of $\sim 3.5 \times 10^{33}$ erg s$^{-1}$, (0.5-7.0 keV; assuming a power-law spectrum with photon index $\Gamma = 1.7$ and an absorbing column density of ${\rm N_H = 6.23 \times 10^{20}\, cm^{-2}}$, based on the average Galactic ${\rm H_{I}}$ column density along the line of sight of this field; \citealt{Dickey90}) at an off-axis angle of $\sim5'$ (Zezas, in prep.). The small number of net counts ($9^{+3}_{-4}$) did not allow us to derive and model the X-ray spectrum for this source.

Despite the low significance of its intensity (1.9$\sigma$ above the background) this is a solid detection (see \citealt{Kashyap10} for a discussion of the detection and intensity significance). This source was also detected by \textit{XMM-Newton} on Dec. 18, 2003 (ObsID 0157960201) and reported in the XMM Serendipitous Source Catalog (3XMM-DR4 Version\footnote{\url{http://xmmssc-www.star.le.ac.uk/Catalogue/3XMM-DR4/}}) as 3XMM J005408.9-724144. We reanalyzed these data with the \textit{XMM-Newton} Science Analysis System (SAS v12.0.1). After processing the raw data with the \textit{epchain} and \textit{emchain} tasks, we filtered any bad columns/pixels and high background flares (excluding times when the total count rate deviated more than 3$\sigma$ from the mean), resulting in 14.8 ks, 18.7 ks, and 17.2 ks net exposures for the European Photon Imaging Camera (EPIC) Metal Oxide Semi-conductor (MOS)1, MOS2, PN cameras, respectively. We only kept events of patterns 0-4 for the PN and 0-12 for the MOS detectors. Source detection was performed simultaneously in five energy bands (0.2-0.5 keV, 0.5-1.0 keV, 1.0-2.0 keV, 2.0-4.5 keV, and 4.5-12.0 keV) for each of the three EPIC detectors with the maximum likelihood method (threshold set to 7) of the \textit{edetect\_chain} task. 
At the position of CXOU J005409.57-724143.5 in the EPIC PN camera, there is source XMMU J005409.2-724143 with coordinates R.A.=00:54:09.16 (J2000.0), Dec.=-72:41:43.46 (J2000.0) and a positional error of 1.6\arcsec (at the 1$\sigma$ level, including the relative as well as the absolute astrometric uncertainty). The \textit{edetect\_chain} task lists this source  with $39\pm9$ counts (source-detection likelihood ${\rm DET\_ML=24.6}$) at an off-axis angle of $4.89'$ in the 0.2-12.0 keV energy band\footnote{Using the revised energy correction factors from \url{http://xmmssc-www.star.le.ac.uk/Catalogue/2XMMi-DR3/UserGuide_xmmcat.html#ProblECFs}.}. The absorption corrected X-ray flux, assuming a spectral model of an absorbed power-law  with a column density ${\rm N_H = 6.23 \times 10^{20}\, cm^{-2}}$ and spectral slope $\Gamma = 1.7$, is then $(1.6 \pm 0.4) \times 10^{-14}\, {\rm erg\, cm^{-2}\, s^{-1}}$. At the distance of the SMC, this corresponds to absorption corrected luminosity of ${\rm L_X^{abs.}=(7.1 \pm 1.7) \times 10^{33}\, erg\, s^{-1}}$ (0.2-12.0 keV). We note that source XMMU J005409.2-724143 was not detected with either of the EPIC MOS detectors, while the XMM Serendipitous Source Catalog lists it with slightly different coordinates and characteristics (${\rm DET\_ML=34.6}$ and mean ${\rm F_{X}\sim(1.0 \pm 0.4) \times 10^{-14}\, erg\, cm^{-2}\, s^{-1}}$), which are consistent with our measurements within the errors. \citet{Novara11} who also analyzed these data do not report this source since they only focus on bright sources. 

We then extracted the PN source spectrum from a 8.5\arcsec radius aperture and a background spectrum from a 75\arcsec source-free region at the same distance from the readout as the source region. 
Unfortunately, the high background in combination with the small number of detected counts did not allow us to perform any spectral analysis. 

The area around source CXOU J005409.57-724143.5 was also observed with \textit{XMM-Newton} on November 01, 2006 (ObsID 0404680201). Following the same analysis procedure as above, we obtain a 30.5 ks net exposure for the EPIC PN camera. Despite the almost double exposure time compared to the 2003 observation, this time the source was undetected. We used the BEHR tool\footnote{\url{http://hea-www.harvard.edu/AstroStat/BEHR/}} \citep{Park06} and measured an intensity upper bound of 10.3 counts (at the 99\% confidence level) at the location of the source. This corresponds to an observed X-ray luminosity of ${\rm L_X^{abs.}}\sim 4.7 \times 10^{32}$ erg s$^{-1}$ (0.2-12 keV) assuming the same spectral parameters and distance as for the other \textit{XMM-Newton} observation.
This is a factor of 10 lower than the previously derived source intensity.

\section{Selection of candidate BeXRBs}
\label{s-select_bexrbs}

Here we use the spectra extracted as described in the previous section in order to identify the spectral types of the studied sources. However, since contamination by the interstellar emission can be significant, for our analysis we selected objects with minimum contamination based on the width of their H$\alpha$ emission line and their [SII]/H$\alpha$ ratio (c.f. \citealt{Antoniou09}). A minimum Full Width at Half-Maximum of the H$\alpha$ emission line (FWHM$_{\rm {H}_\alpha}$) is used to eliminate objects with too narrow emission, since BeXRBs have broader emission lines than the interstellar component \citep{Coe05}. The selection of objects with [SII]/H$\alpha$ ratio smaller than in the average "sky" spectrum also helps to eliminate objects with contamination from diffuse interstellar emission (mainly supernova remnants).

The minimum FWHM$_{\rm {H}_\alpha}$ was based on the width of the H$\alpha$ line measured in the sky spectra separately for each observation as the use of different gratings resulted in different spectral resolutions. For the July 26 data, there are only 3 sky spectra of suitable S/N (two from the north and one from the south field) which result in an average interstellar H$\alpha$ width of $<$FWHM$_{\rm {H}_\alpha}>=5.70\pm0.07$ {\AA}. For the September 19 data, 9 good sky spectra were used resulting in an interstellar H$\alpha$ width of $<$FWHM$_{\rm {H}_\alpha}>=1.94\pm0.03$ {\AA}. 
We selected for further analysis objects with a FWHM$_{\rm {H}_\alpha}$ at least 3$\sigma$ above the average width of the interstellar H$\alpha$ emission. This means that only objects with FWHM$_{\rm {H}_\alpha}$\textgreater 5.91 {\AA} for July 26,  and FWHM$_{\rm {H}_\alpha}$\textgreater 2.03 {\AA} for September 19, were considered for further examination.  

In order to determine the average [SII]/H$\alpha$ ratio for the interstellar emission we used again the sky spectra.
We found [SII]/H$\alpha$=0.28$\pm$0.11 for the south field (September 19) and [SII]/H$\alpha$=0.38$\pm$0.02 for the north field (July 26). As no H$\alpha$ emission was present in the only sky spectrum of the south field observed on July 26, we used the [SII]/H$\alpha$ ratio measured in the September 19 run (since the flux of the lines is independent of the spectral resolution). These values are in agreement with the [SII]/H$\alpha$ ratio ($\geq$0.4) expected in environments with supernova remnants, indicating that the ISM in these regions is to a large degree shock excited. Thus, all sources considered as BeXRB candidates should have a [SII]/H$\alpha$ ratio smaller than the maximum values found for each field, in addition to the low limit on the FWHM$_{\rm {H}_\alpha}$. 
 
After applying these criteria, we are left with 21 sources out of the 272 initial targets that are BeXRB candidates and which are considered for spectral classification. These include 18 \textit{Chandra} and 3 \textit{XMM-Newton} sources.  

The final spectra were normalized by subtracting the stellar continuum (after a spline fit), using the DIPSO v3.6-3 package of STARLINK \citep{Howarth04}.

\section{Spectral classification}
\label{s-spectral_class}

\begin{table}
\caption{Classification criteria for B-type stars in SMC from Antoniou et al. (2009) and Evans et al. (2004).
}
 \begin{tabular}{lc}
 \hline\hline
  Line identifications & Spectral Type \\
 \hline
  HeII $\lambda$4200,HeII $\lambda$4541,HeII $\lambda$4686 present & earlier than B0 \\
  HeII $\lambda$4541 and HeII $\lambda$4686 present, HeII $\lambda$4200 weak & B0 \\
  HeII $\lambda$4200 and HeII $\lambda$4541 absent, HeII $\lambda$4686 weak & B0.5 \\
  HeII $\lambda$4686 absent, SiIV $\lambda\lambda$4088,4116 present & B1 \\
  SiIV $\lambda$4116 absent, SiIII $\lambda$4553 appear & B1.5 \\
  OII+CIII $\lambda$4640-4650 blend decreases rapidly & later than B1.5 \\
  SiIV and SiII absent, MgII $\lambda4481 \textless$  SiIII $\lambda$4553 & B2 \\ 
  MgII $\lambda4481 \sim$ SiIII $\lambda$4553 & B2.5 \\
  MgII $\lambda4481 \textgreater$ SiIII $\lambda$4553 & B3 \\
  OII+CIII $\lambda$4640-4650 blend disappears, & later than B3 \\
  OII $\lambda$4415-4417, NII $\lambda$4631 disappear & \\
  clear presence of HeI $\lambda4471$ and absence of MgII $\lambda$4481 & earlier than B5\\
  SiIII $\lambda$4553 absent, SiII $\lambda4128-4132 \textless$ HeI $\lambda$4121,  & B5 \\
  HeI $\lambda4121 \textless$ SiII $\lambda4128-4132 \textless$ HeI $\lambda$4144, & B8 \\
  MgII $\lambda4481 \leqslant$ HeI $\lambda$4471 &   \\
  HeI $\lambda4471 \textless$ MgII $\lambda$4481, & B9 \\
  FeII $\lambda4233 \textless$ SiII $\lambda$4128-4132 &  \\
 \hline
 \end{tabular}
\label{t-classification_criteria}
\end{table}

\begin{figure*} 
 \includegraphics[scale=0.59]{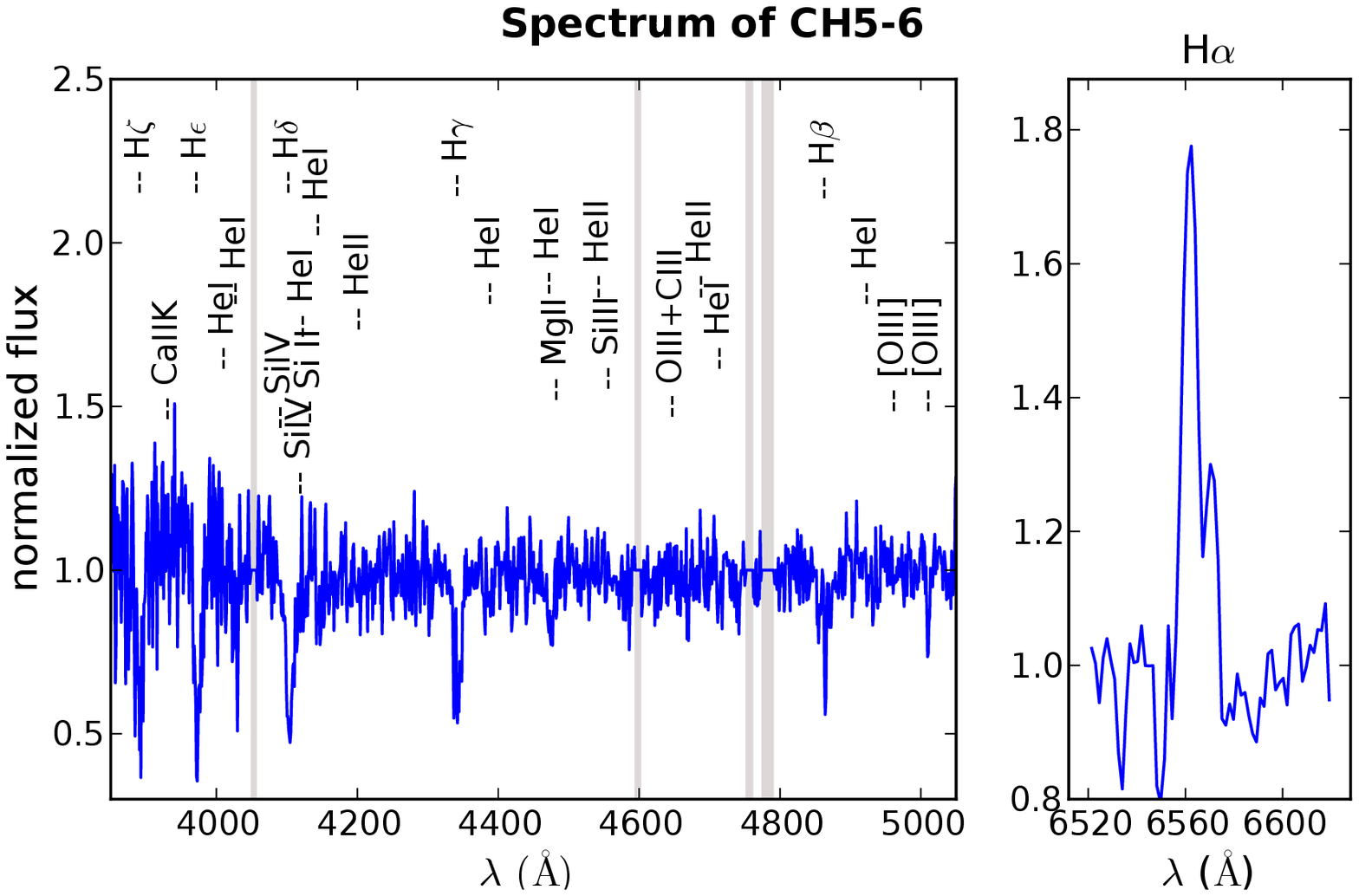} \\
 \includegraphics[scale=0.59]{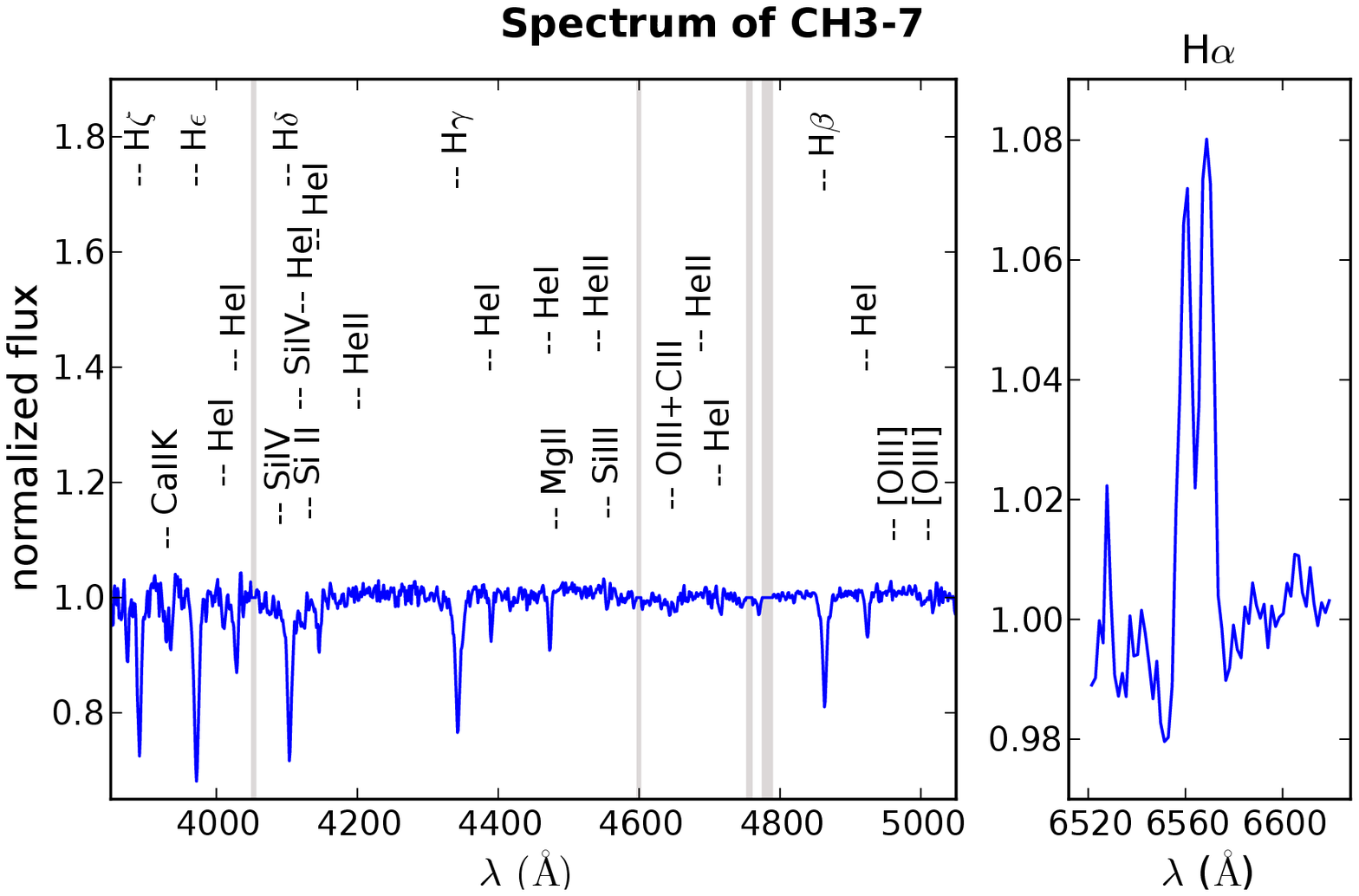} \\
 \includegraphics[scale=0.59]{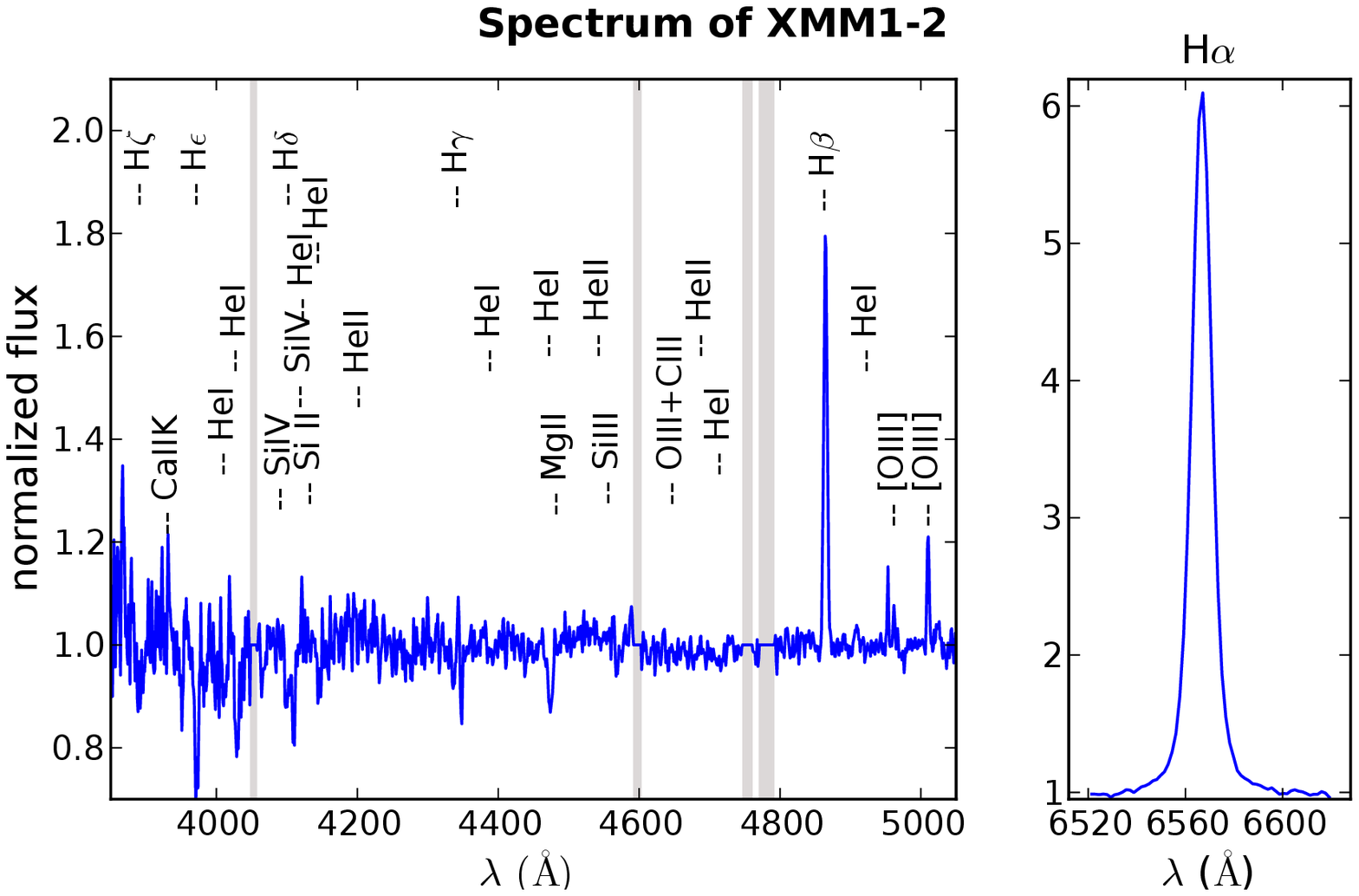}  
 \caption{The spectra of the 5 new BeXRBs identified in this work (including the tentative BeXRB CH7-19). Shaded areas indicate wavelength ranges of bad columns and/or sky subtraction residuals.}
 \label{f-new_spectra}
\end{figure*}

\begin{figure*}
 \includegraphics[scale=0.59]{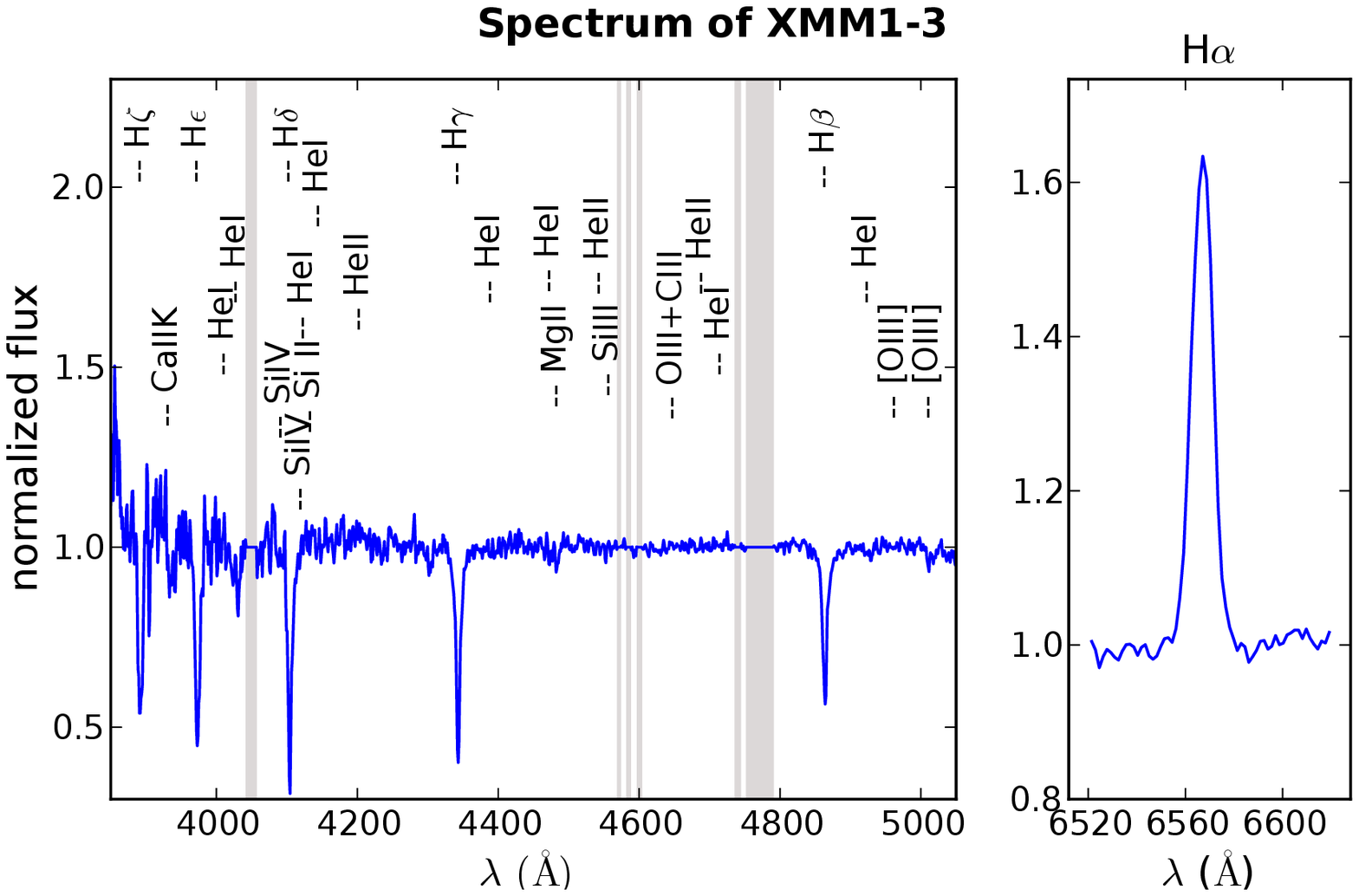} \\  
 \includegraphics[scale=0.59]{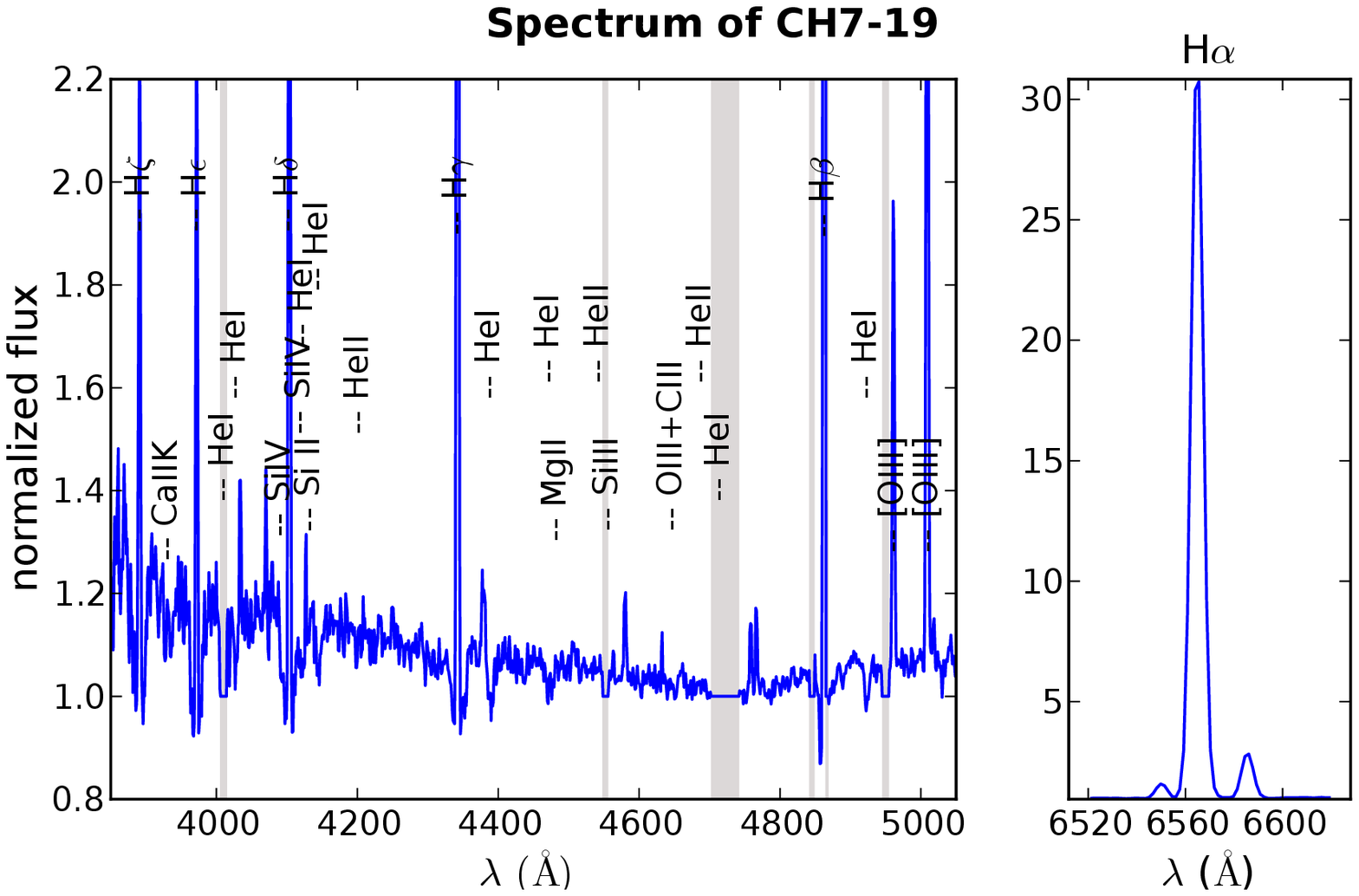} \\
 \contcaption{}
\end{figure*}

\begin{figure*}
 \includegraphics[scale=0.59]{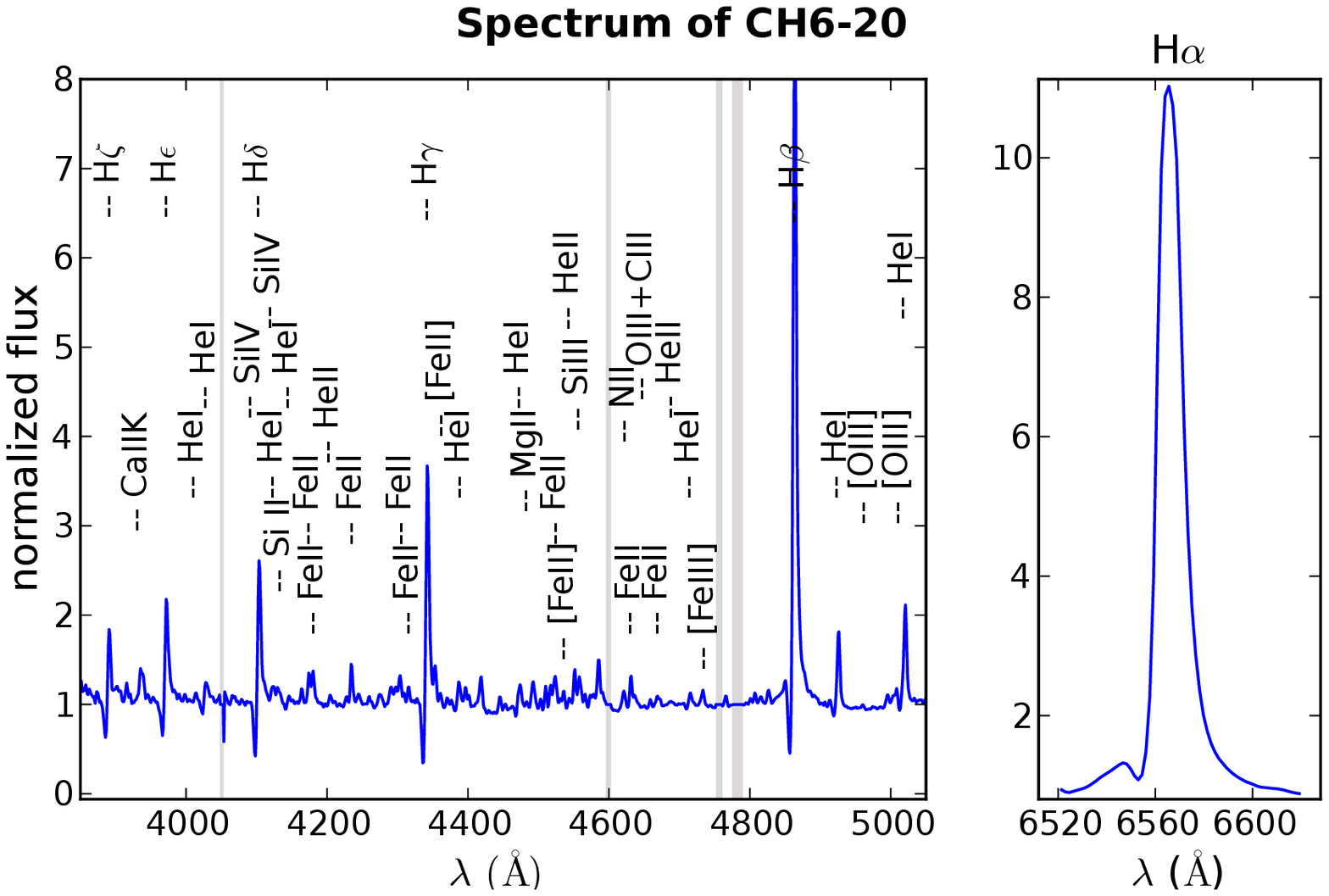} \\
 \includegraphics[scale=0.59]{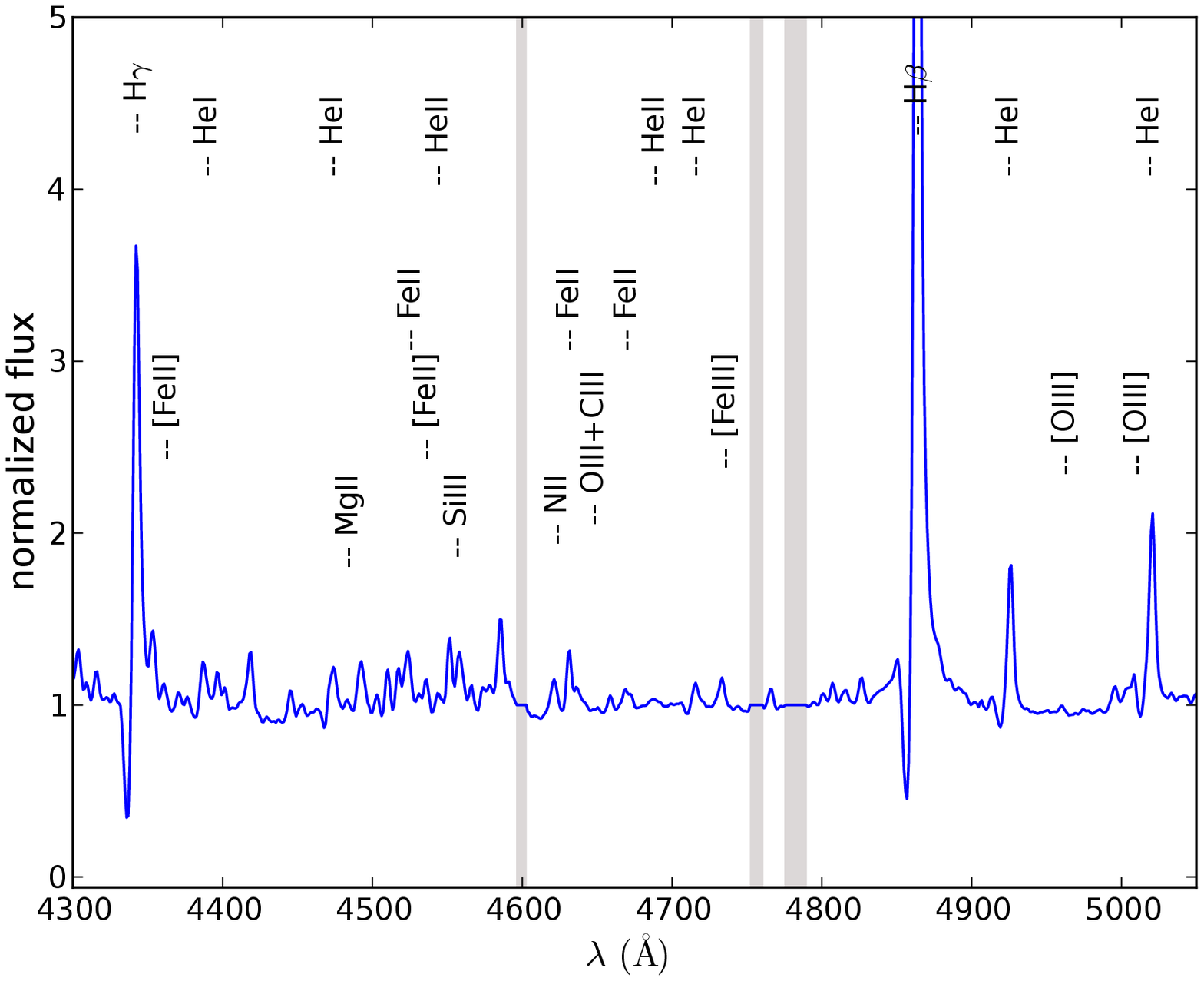}
 \caption{Spectrum of the optical counterpart to the \textit{Chandra} source CXOU J005409.57-724143.5 (CH6-20) identified as a sgB0[e] star \citep{Zickgraf89}, which is the known sgB[e] star LHA 115-S 18 \citep{Henize56}. There is clear presence of emission lines of HeII, permitted  and forbidden Fe lines, and Balmer lines with P Cygni profiles.}
 \label{f-sg_spectrum}
\end{figure*}

\subsection{Spectral classification criteria}
\label{s-spectral_class_criteria}

As seen in the previous section, the selection of sources for further classification was based on their broad H$\alpha$ lines, a key characteristic of BeXRBs. In Figs. \ref{f-new_spectra}, \ref{f-sg_spectrum}, and \ref{f-extra_spectra}, we present the spectra of the 21 sources selected for further analysis. As expected, by selection, they exhibit strong H$\alpha$ emission. The left-hand panels in these figures show the blue part of the spectrum with the diagnostic lines for spectral-type classification marked, while the right-hand panels focus on the H$\alpha$ line. Shaded areas indicate bad columns and/or sky subtraction residuals. We see that two sources (CH4-8, and CH4-2 in Fig. \ref{f-extra_spectra}) show asymmetric H$\alpha$ profiles, while two more sources (CH3-7, and CH5-6 in Fig. \ref{f-new_spectra}), show double-peaked H$\alpha$ emission. Although Herbig Ae/Be stars present most of the time a double-peaked H$\alpha$ profile \citep{Vieira03}, we can safely rule out this possibility since the position of our objects in the \textit{V, B-V} CMD (see fig. 3 of \citealt{Antoniou09bb}) ensures that they are not pre-main sequence objects. One more source (CH6-20) shows evidence for P-Cygni profiles (Fig. \ref{f-sg_spectrum}). 

The spectral-type classification is based on the scheme of \citet{Evans04}, which was later applied to BeXRBs by \citet{McBride08} and \citet{Antoniou09}, supplemented by the atlas of \citet{Gray09}. In order to classify OB stars, the use of metal lines is normally preferred, but because of the low metallicity of the SMC they are much weaker and difficult to detect. Thus, we classify the spectra based on a combination of HeI, HeII, and metal lines.
The spectral lines used for our classification are presented in Table \ref{t-classification_criteria}.

In Table \ref{t-source_list} we present the list of the HMXBs identified in our study, along with the classification derived from this and previous studies. In this table each source is identified with an ID of the type CH/XMM F-NN, where CH stands for \textit{Chandra} and XMM for \textit{XMM-Newton} sources, F is the field number, and NN is the source ID from the studies of \citet{Antoniou09bb}, and \citet{Antoniou10}, respectively. We find 20 sources of B spectral type which in combination with their strong H$\alpha$ lines and X-ray emission make them BeXRBs, and one X-ray emitting supergiant system with also strong emission in the H$\alpha$ line and evidence for a strong wind, which makes it a possible HMXB. Next we discuss the new BeXRBs and the supergiant system identified in this work; sources that have been already classified as BeXRBs are discussed in detail in the Appendix. 

For each source we measured the equivalent width ($EW$) of the H$\alpha$ line as the ratio of the flux in the region of H$\alpha$ (6553-6576 {\AA}) over the continuum at the position of H$\alpha$ as calculated by a linear fit to the flux of line-free regions adjacent to H$\alpha$ (i.e. at 6530-6540.5 {\AA} and 6630-6650 {\AA}). These measurements are presented in Table \ref{t-ha_ew_list}.

\subsection{Discussion of individual sources}
\label{s-discus_sources}

\begin{itemize}
 \item \textit{CXOU J005409.57-724143.5 (source CH6-20) - classified as sgB0[e]}

This is probably the most interesting object in our sample (see Fig. \ref{f-sg_spectrum}). Not only all the Balmer lines are in emission but we also observe HeII, as well as permitted and forbidden Fe lines in emission. Moreover, all the Balmer lines and some HeII lines present P Cygni profiles, typical of supergiant B[e] stars \citep{Shore87,Zickgraf89,Lamers98}. However, due to the lack of characteristic lines (which are absent even in high resolution spectra, c.f. \citealt{Graus12}), we cannot identify the spectral type of this source.
Through a model atmosphere fit \citet{Zickgraf89} concluded that the stellar temperature corresponds to a B0 spectral type. According to the classification criteria of Table \ref{t-classification_criteria} the spectrum should also show HeII $\lambda$4686 for this spectral type, which is however absent in our spectrum. Nevertheless, the star is known to show variability in this line \citep{Shore87,Morris96}. To the best of our knowledge, there are no other classifications in the literature, thus we classify source CH6-20 as a sgB0[e]. This makes this source the second X-ray source associated with a supergiant in the SMC. We further discuss the nature of source CH6-20 in Section \ref{s-sgcase}.   

 \item \textit{CXOU J005504.40-722230.4 (source CH5-6) - classified as B1-B5}
 
The HeII  $\lambda\lambda$4200, 4686 lines are absent so this spectrum is of B1 spectral type or later. As the HeI $\lambda$4471 line is present without any sign of the MgII $\lambda$4481 line, we can deduce that the source must be earlier than B5. No other line that would help us to further constrain the classification is visible, thus we only propose a B1-B5 spectral-type range for source CH5-6. 

 \item \textit{CXOU J005723.77-722357.0 (source CH3-7) - classified as B2}

The HeI $\lambda\lambda$4144, 4387, 4471 absorption lines are clearly detected. Since no HeII $\lambda\lambda$4200, 4686 lines are visible the source has a spectral type of B1 or later. Moreover, the absence of the MgII $\lambda$4481 line and the presence of SiIII indicates that the star is earlier than B2.5, which is also supported by the presence of the OII $\lambda$4415-4417 lines and the OII+CIII $\lambda$4640-4650 blend. The SiIV $\lambda\lambda$4116, 4088 and SiII $\lambda$4128-4132 lines are not present, which means that all criteria for a B2 star are completely fulfilled for this previously unclassified source. 

 \item \textit{XMMU J010519.9-724943 (source XMM1-2) - classified as B3-B5}
 
The absence of the HeII $\lambda\lambda$4200, 4686 lines supports a spectral type of B1 or later. We see no evidence for the OII+CIII $\lambda$4640-4650 blend, so the source must be later than B3, as this blend disappears after this class. Moreover, this source cannot be later than B5, since the HeI $\lambda$4471 line is present but without any sign of the MgII $\lambda$4481 line. We thus assign to the previously unclassified source XMM1-2 a spectral type of B3-B5. Noteworthy are the weak H$\gamma$ emission and the almost filled-in (by emission) H$\delta$ line.

 \item \textit{XMMU J010620.0-724049 (source XMM1-3) - classified as B9}
 
The strong presence of the MgII $\lambda$4481 line is characteristic of a late B-type star. Moreover, the MgII $\lambda$4481 line is clearly stronger than the HeI $\lambda$4471 line, which immediately places this source in the spectral range later than B8. The CaII K $\lambda$3933 line is also present in later B types and it becomes a dominant metal line in A-type stars (being much stronger than MgII). In our case the CaII K $\lambda$3933 line is much stronger than the usual interstellar emission line seen in all other spectra, but it is weaker than the MgII $\lambda$4481 line, so we conclude that this, previously unclassified, star has a B9 spectral type. 

 \item \textit{CXOU J004941.43-724843.8 (source CH7-19) - classified as B1-B5}
 
This source displays an H$\alpha$ emission line with a FWHM$_{\rm {H}_\alpha}=5.21$ {\AA} which is marginally lower than our limiting H$\alpha$ FWHM (FWHM$_{\rm {H}_\alpha}=5.70\pm0.07$ {\AA} for July 26; c.f. Section \ref{s-select_bexrbs}). Nevertheless, we decided to further examine this source. The most striking features of its spectrum are the emission in the Balmer lines series and the presence of the  [OIII] $\lambda\lambda$4959, 5007 lines. As the sky subtraction is not perfect the [OIII] lines are residuals of the contribution from the surrounding environment. A previous classification for this source has been given by \citet{Murphy00} as a potential Planetary Nebula (source 61 in their list). Their analysis showed that only 33\% of the good candidates are real Planetary Nebulae, while source 61 is not characterized as a good candidate as its properties are closer to emission stars. Taking into account that their position accuracy was not better than 12" and that there is a star cluster in the same region \citep{Bica95}, it is possible that this source was misclassified as a potential Planetary Nebula. In addition, the FWHM$_{\rm {H}_\alpha}=5.21$ {\AA} translates to a rotational velocity of ${\rm vsini}\sim$240 km s$^{-1}$ which is within the normal range of rotational velocities for Be stars \citep{Steele99}. Thus, the major contributor of the spectral lines is considered to be of stellar nature than interstellar.
 The absence of both HeII $\lambda\lambda$4200, 4686 lines suggests that the source is of spectral type B1 or later, but not later than B5 due to the absence of the MgII $\lambda$4481 line. Thus, we tentatively classify the previously unclassified source CH7-19 as B1-B5, although the Be nature is uncertain due to the marginal width of its H$\alpha$ emission line. 

\end{itemize}

\section{Discussion}
\label{s-discussion}

\subsection{New Be/X-ray Binaries}
\label{s-new_bexrbs}

In Table \ref{t-source_list} we present the X-ray and optical properties of the BeXRBs studied in this work. From the 21 classified sources, 12 are in full agreement (within 0.5 spectral type) with the previous studies of \citet{McBride08}, and \citet{Antoniou09}, while 3 have later spectral types (by 1 to 1.5 type) than previous results. Most importantly, we identify 4 new BeXRBs and 1 candidate sgXRB system (discussed in detail in Section \ref{s-sgcase}). Although source CH7-19 is included in the table we exclude this source from further analysis, as its identification as an emission line star is tentative due to the possibly significant contamination of its spectrum by interstellar emission (see Sections \ref{s-select_bexrbs} and \ref{s-discus_sources}). In the following discussion, we exclude from any comparisons with other samples source CH6-20 (the sgB[e]) since it does not belong to the same population as the luminosity class III-V BeXRBs. The remaining BeXRBs is hereafter referred to as the "sample".    

The 4 new BeXRBs sources (hereafter referred to as "new") are: CXOU J005504.40-722230.4 (CH5-6), CXOU J005723.77-722357.0 (CH3-7), XMMU J010519.9-724943 (XMM1-2), and XMMU J010620.0-724049 (XMM1-3). The spectral types for sources CH3-7 and XMM1-3 (B2 and B9, respectively) are accurate to $\pm$0.5 subclass. XMM1-3 is the BeXRB with the latest spectral type (B9) known in the SMC. For the remaining new BeXRBs, we can provide only a range of spectral types: B1-B5 for the source CH5-6, and B3-B5 for the source XMM1-2. 

All but source CH3-7 (\textit{V}=14.71 mag), are faint objects with \textit{V}-magnitudes in the range 16.4-17.9 mag (see Table \ref{t-source_list}) and have rather noisy spectra (see Fig. \ref{f-new_spectra}) which hampers their spectral classification.
They are also faint X-ray sources with typical X-ray luminosities $\sim10^{34}$ erg s$^{-1}$, outside the typical range of luminosities of outbursting BeXRBs. Overall our sample of BeXRBs spans 3 orders of magnitude in X-ray luminosity and it includes faint BeXRBs for which it is not possible to detect X-ray pulsations.

\subsection{Spectral-type distributions}
\label{s-spectral_distributions}

\begin{figure}
\resizebox{\hsize}{!}{\includegraphics{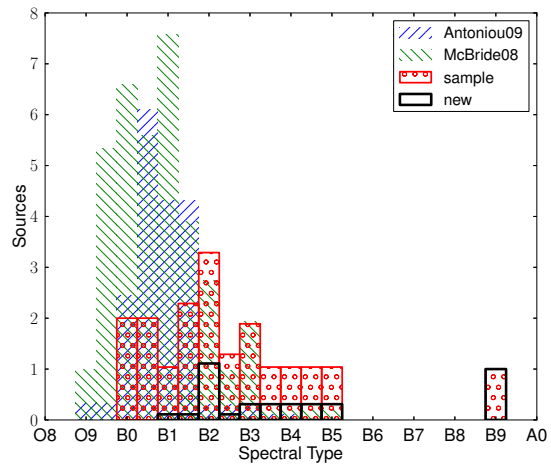}}
\caption{Comparison of spectral-type distributions of the different samples considered in this work: "Antoniou09" (blue, right diagonal line filled) and "McBride08" (green, left diagonal line filled) corresponds to the spectral-type distributions obtained from \citet{Antoniou09} and \citet{McBride08}, respectively; "sample" (red solid line, 'o' filled) and "new" (black solid line) correspond to the BeXRB sample studied in this work (excluding sources CH7-19 and CH6-20) and only the new sources (excluding source CH6-20) from this work respectively, as defined in Section \ref{s-new_bexrbs}.
(For sources extending over more than one class their spectral type is split equally between the encompassed class bins, e.g. a B0-B2 object will split into 1/3 in B0, B1, and B2 spectral type, respectively.)
}
\label{f-spectral_distributions}
\end{figure}

\begin{figure}
\resizebox{\hsize}{!}{\includegraphics{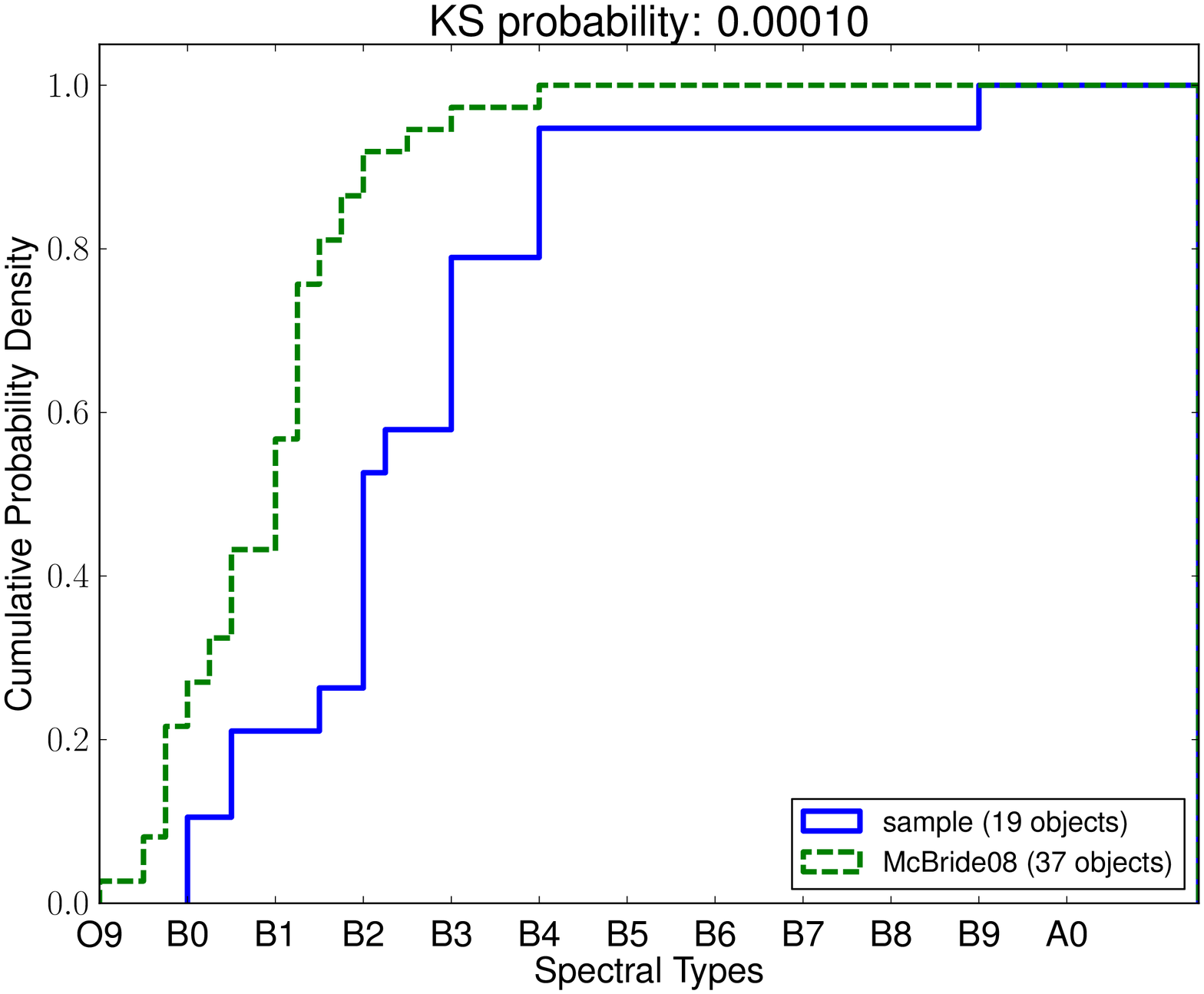}} 
\resizebox{\hsize}{!}{\includegraphics{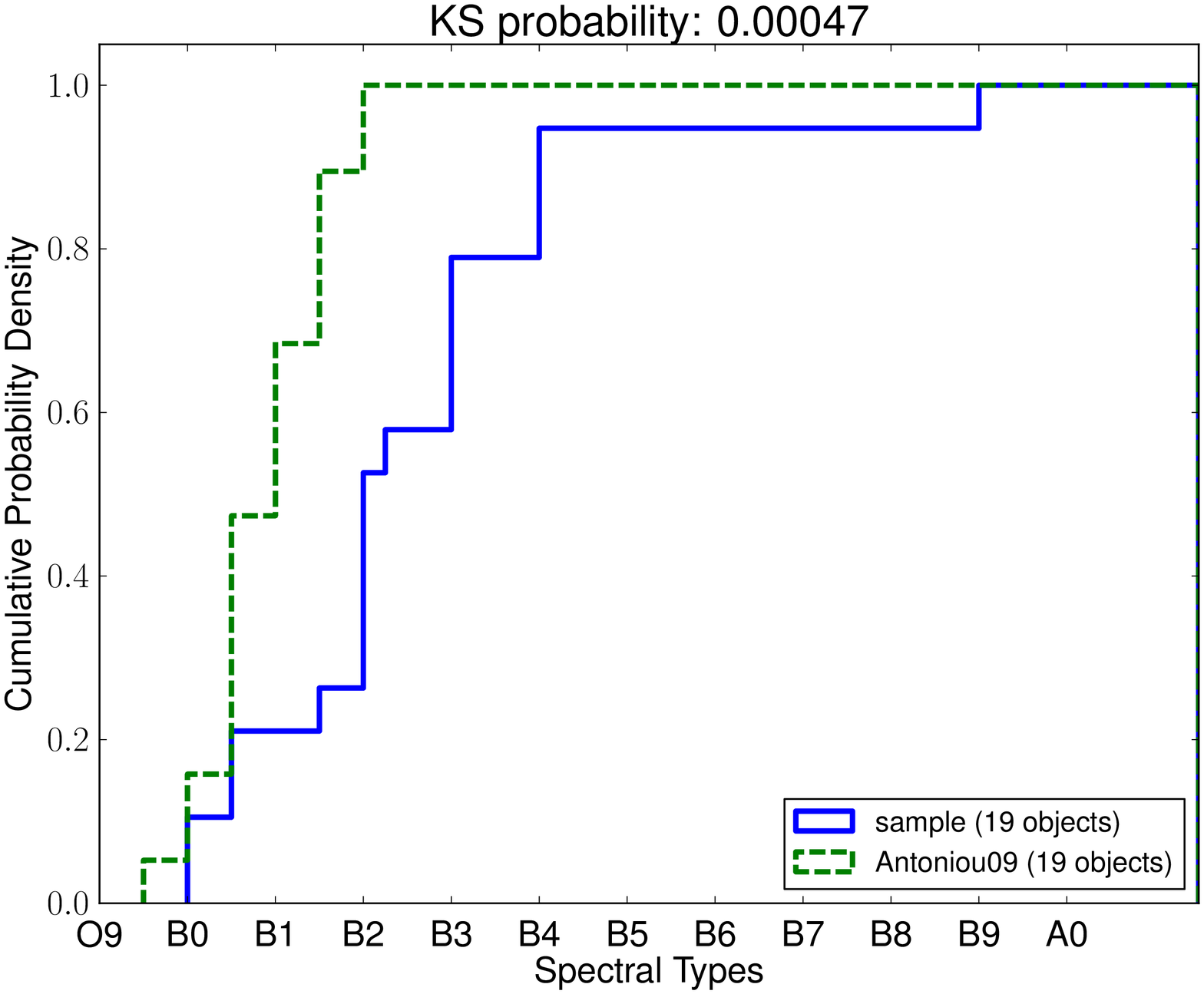}}
\caption{Cumulative distributions of the spectral types of BeXRBs populations in the SMC for our sample (solid blue, excluding source CH6-20) compared with the samples (dashed green) of \citet{Antoniou09} and \citet{McBride08}. By applying the Kolmogorov-Smirnov test we find that our sample is different from the previous ones, at more than 99\% confidence level (the probability to reject the null hypothesis, that the two distributions come from the same parent distribution, is given above each plot). This indicates that our sample is skewed to later spectral types. 
}
\label{f-kstest_spectral_distributions}
\end{figure}

By combining our results with these from previous studies of the BeXRBs in the SMC (e.g. \citealt{McBride08,Antoniou09}) we can obtain a more complete sample of the properties of the BeXRB population in the SMC. 
In Fig. \ref{f-spectral_distributions} we plot a histogram of the spectral-type distribution of our sample and new sources separately (as defined in Section \ref{s-new_bexrbs}), along with the samples of the previous studies by \citet{McBride08} and \citet{Antoniou09}. In order to account for the uncertainty in the spectral-type classification we split sources extending over more than one class equally between the encompassed class bins (e.g. a B0-B2 object will split into 1/3 in B0, B1, and B2 spectral class, respectively.)
By inspecting this figure we see that there is a trend for our sample to extend to later spectral types than previous works. In order to assess the significance of this trend we compared the spectral-type distributions of these samples using a two-sample Kolmogorov-Smirnov test \citep[KS test;][]{Conover99}.
 We find that this trend is significant at more than 99\% confidence level, which further indicates that the sources in our sample are skewed towards later types (see Fig. \ref{f-kstest_spectral_distributions} for the cumulative distributions of the spectral types in these three samples). 

\begin{figure}
\resizebox{\hsize}{!}{\includegraphics{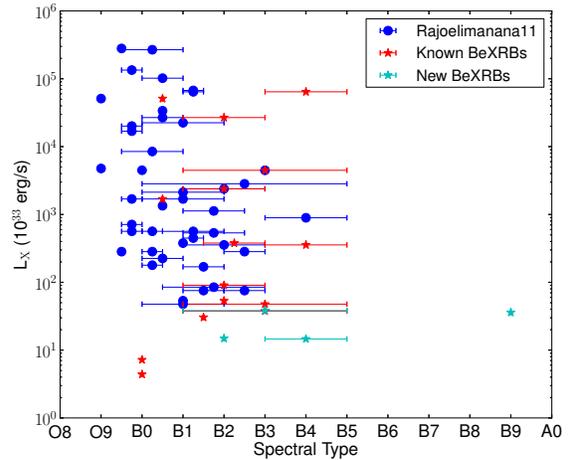}}
\caption{Unabsorbed X-ray luminosities (${\rm L_X}$) of SMC BeXRBs plotted against the spectral type of their optical counterparts. The blue circles  represent the maximum observed ${\rm L_X}$ for sources obtained from table 1 of \citet{Rajoelimanana11} and after transforming these values to the 0.5-7 keV energy band that we used in our study (assuming a power law with photon index $\Gamma=1.7$ and $\rm {N_H} \sim 6 \times 10^{20}$  cm$^{-2}$). The red asterisks ("Known BeXRBs") represent sources from our sample (see Table \ref{t-source_list}) with a previous classification and their maximum luminosities are derived either from table 1 of \citet{Rajoelimanana11} or from Zezas (in prep., \textit{Chandra} sources) and \citet[][\textit{XMM-Newton} sources]{Antoniou10}. The cyan asterisks ("New BeXRBs") represent new sources identified in this work (see Section \ref{s-new_bexrbs}) with their luminosities taken from Zezas (in prep.), and \citet{Antoniou10}. 
}
\label{f-specdis-Lx}
\end{figure}

In Fig. \ref{f-specdis-Lx} we also present the unabsorbed X-ray luminosities of the sources identified in this study (new identifications are presented again separately) along with the maximum X-ray luminosity of known BeXRBs in the SMC, as given in table 1 of \citet{Rajoelimanana11}. As this list is not homogeneous we transformed these values to the 0.5-7 keV energy band that we used in our study, assuming a power law with photon index $\Gamma=1.7$ and $\rm {N_H}= 6 \times 10^{20}$ cm$^{-2}$. We obtained the maximum luminosities for our sources from table 1 of \citet{Rajoelimanana11}. For the sources that are not in this catalog we used the unabsorbed luminosities presented in Zezas (in prep.) and \citet{Antoniou09bb,Antoniou10}. We have also assumed that the X-ray luminosities given in \citet{Rajoelimanana11} are unabsorbed. In any case, the difference between the unabsorbed and absorbed luminosities for the assumed model and energy band is not important for the purpose of the comparison presented here. 

In  Fig. \ref{f-specdis-Lx} we see a weak trend for lower-luminosity sources to be associated with sources of wider spectral-type range, and extending to later types.   
A possible physical explanation for this trend may lie in the nature of the BeXRBs, especially if there is a correlation between the spectral type and the size of the equatorial disk. 
Although this is a very intriguing prospect, we should note that the observed trend could be the result of a selection bias: the list of \citet{Rajoelimanana11} gives by construction the historically maximum detected X-ray luminosity for these sources, 
while our sample, which mainly contributes to the lower luminosity, later spectral-type sources, includes sources for which the X-ray luminosity was measured from a single snapshot.

\begin{figure}
\resizebox{\hsize}{!}{\includegraphics{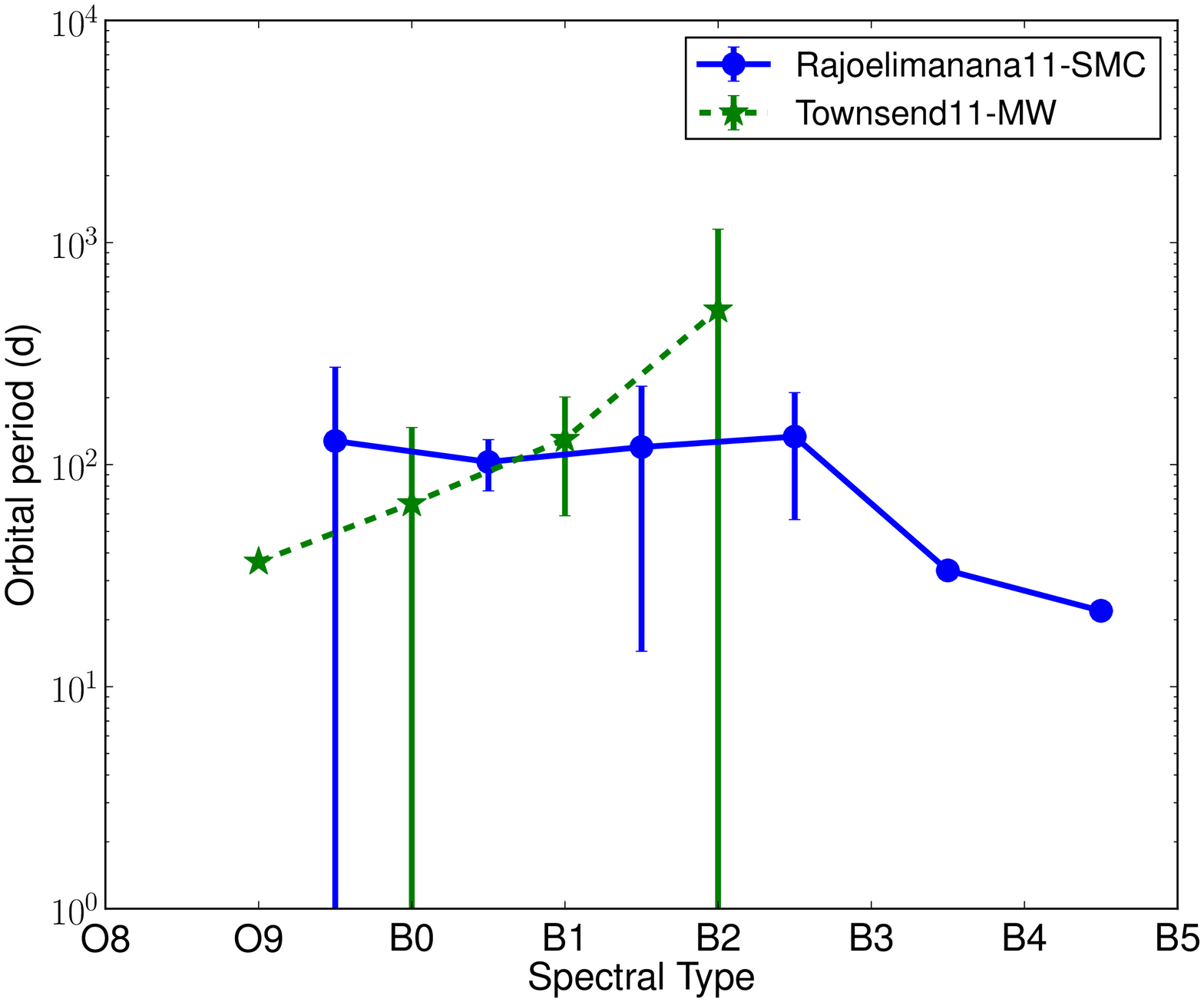}}
\resizebox{\hsize}{!}{\includegraphics{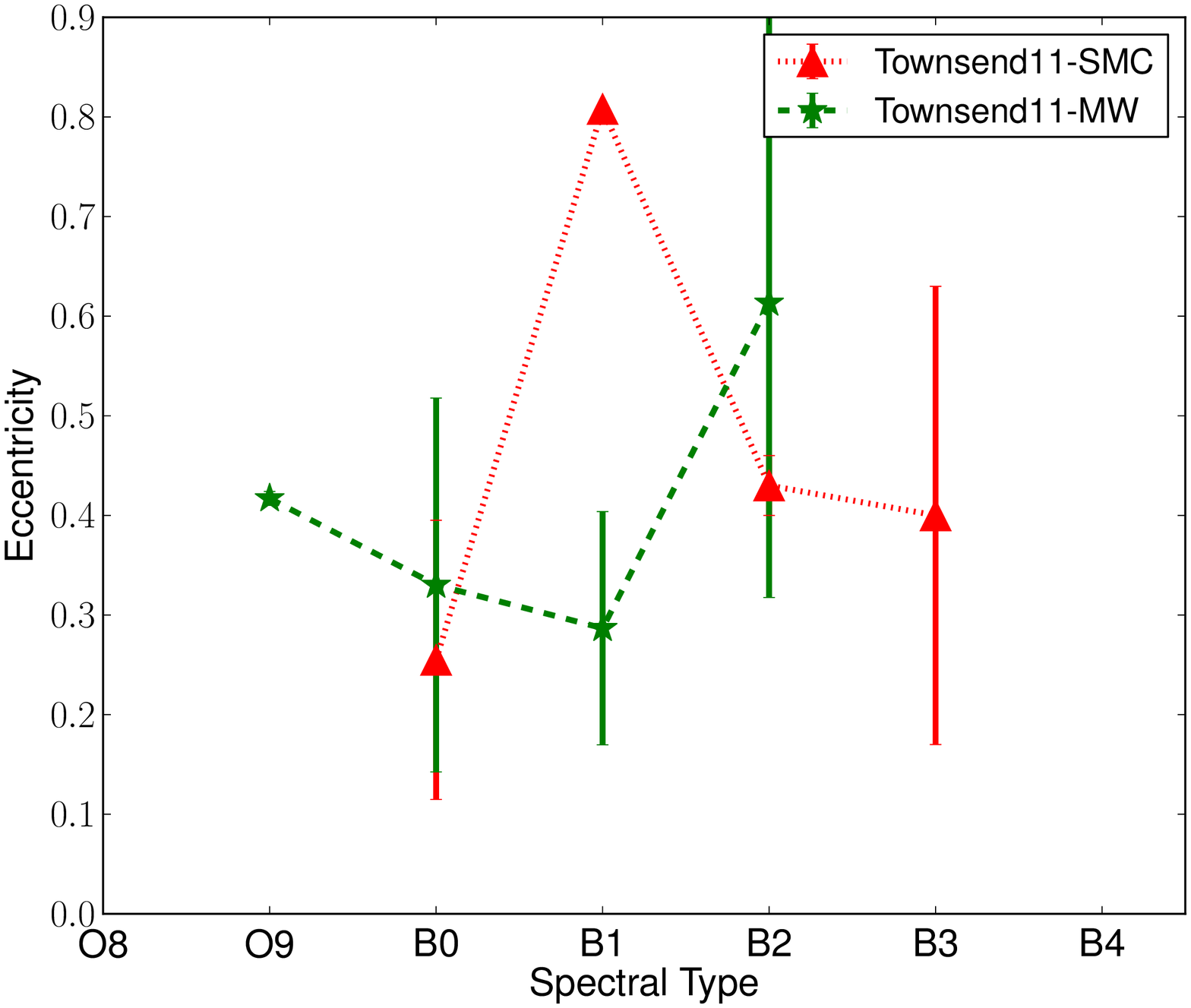}}
\caption{The orbital periods ($Porb$) and eccentricities ($e$) as a function of the spectral types of the optical counterparts of BeXRBs populations in the SMC and the Milky Way (binned to 1 spectral type). For the orbital periods the solid blue and the dashed green lines correspond to the samples of BeXRBs in the SMC (after \citealt{Rajoelimanana11}) and in the Milky Way respectively (after \citealt{Townsend11}).
For the eccentricities we use data from \citet{Townsend11} for BeXRBs in the SMC (dotted red) and Milky Way (dashed green).  The error bars indicate the $1 \sigma$ standard deviation for each of the two parameters within each spectral-type bin. 
}
\label{f-specdis-orbital_param}
\end{figure}

In Fig. \ref{f-specdis-orbital_param} we compare the average orbital period ($Porb$) and eccentricity ($e$) as a function of spectral type of the BeXRB populations in the SMC and Milky Way (MW). Data for the orbital periods of BeXRBs in the SMC are taken from \citet{Rajoelimanana11}, and for the MW from \citet{Townsend11}. For the eccentricities we take all data (for both SMC and MW) from \citet{Townsend11}.

{
\setlength{\tabcolsep}{4.5pt}
\begin{landscape}
 \begin{table}
  \begin{minipage}{\linewidth} 
  \caption{Optical and X-ray properties of the studied sources.}
   \begin{tabular}{ l | c c c c c c | c c c c | l l }
   \hline\hline
  X-ray & Optical  & \multicolumn{5}{c}{Optical Counterpart} & \multicolumn{4}{c}{X-ray Source} & \multicolumn{2}{c}{Classification}\\
 source ID & source ID & RA & Dec & Offset & \textit{V} & \textit{B-V} & ID & RA & Dec & ${\rm L_{X,unabs}}$ & this work & previous\\
           &    &\multicolumn{2}{c}{(J2000)} & &  & & CXOU=C, XMMU=X & \multicolumn{2}{c}{(J2000)} &  & & \\ 
    &    & (h m s) & ($^o$ ' ") & (") & (mag) & (mag) & & (h m s) & ($^o$ ' ") & ($10^{33} $ erg s$^{-1}$) & & \\ 
(1) & (2) & (3) & (4) & (5) & (6) & (7) & (8) & (9) & (10) & (11) & (12) & (13) \\ 
\hline
  CH4-8 & O\_4\_171264 & 00 48 14.13 & -73 10 03.5 & 0.63 & 15.74$\pm$0.04 & 0.00$\pm$0.05 & C J004814.15-731004.1 & 00 48 14.15 & -73 10 04.1 & 30.4 & B1.5 & B1.5 [A09] \\
  CH7-1 & O\_5\_65517 & 00 49 03.34 & -72 50 52.1 & 0.45 & 16.94$\pm$0.06 & 0.09$\pm$0.10 & C J004903.37-725052.5 & 00 49 03.37 & -72 50 52.5 & 79.2 & B1-B5 & $\sim$B3 [M08] \\ 
  CH4-2 & O\_5\_111490 & 00 49 13.63 & -73 11 37.4 & 0.47 & 16.52$\pm$0.02 & 0.10$\pm$0.04 & C J004913.57-731137.8 & 00 49 13.57 & -73 11 37.8 & 76.6 & B3-B5 & B1.5	[A09]\\
  CH4-5 & O\_5\_111500 & 00 49 29.81 & -73 10 58.0 & 0.61 & 16.30$\pm$0.01 & 0.09$\pm$0.02 & C J004929.74-731058.5 & 00 49 29.74 & -73 10 58.5 & 37.6 & B1-B5 & B1 [A09]	\\ 
  CH7-19 & O\_5\_146766 & 00 49 41.66 & -72 48 42.9 & 1.36 & 17.16$\pm$0.55 & 0.27$\pm$0.60 & C J004941.43-724843.8 & 00 49 41.43 & -72 48 43.8 & 3.7 & B1-B5* & unclassified\\
  CH4-3 & O\_5\_271074 & 00 50 57.12 & -73 10 07.7 & 0.28 & 14.54$\pm$0.01 & -0.06$\pm$0.01 & C J005057.16-731007.9 & 00 50 57.16 & -73 10 07.9 & 90.4 & B1-B5 & B0.5	[A09]\\ 
  CH5-3 & O\_6\_85614 & 00 51 53.11 & -72 31 48.3 & 0.54 & 14.90$\pm$0.12 & -0.27$\pm$0.13 & C J005153.16-723148.8 & 00 51 53.16 & -72 31 48.8 & 57.1 & B0.5 & O9.5-B0 [M08]	\\  
  CH5-1 & Z\_2311496 & 00 52 05.69 & -72 26 04.0 & 0.55 & 14.91$\pm$0.02 & 0.00$\pm$0.03 & C J005205.61-722604.4 & 00 52 05.61 & -72 26 04.4 & 1093.9 & B3-B5 & B1-1.5 [M08]\\
  CH6-1 & O\_6\_77228 & 00 52 08.95 & -72 38 02.9 & 0.58 & 15.03$\pm$0.02 & 0.14$\pm$0.03 & C J005208.95-723803.5 & 00 52 08.95 & -72 38 03.5 & 2342.0 & B1-B5 & B1-3 [A09] \\
  CH5-12 & Z\_2406014 & 00 52 45.10 & -72 28 43.4 & 0.35 & 14.92$\pm$0.08 & 0.00$\pm$0.09 & C J005245.04-722843.6 & 00 52 45.04 & -72 28 43.6 & 7.2 & B0 & O9-B0 [A09] \\
  XMM2-1 & Z\_2430066 & 00 52 55.27 & -71 58 06.0 & 2.82 & 15.53$\pm$0.02 & -0.05$\pm$0.04 & X J005255.1-715809 & 00 52 55.10 & -71 58 08.7 & 893.2 & B1-B3 & B0-B1 [M08]\\ 
  CH5-16 & Z\_2573354 & 00 53 55.38 & -72 26 45.3 & 0.83 & 14.72$\pm$0.03 & -0.07$\pm$0.03 & C J005355.25-722645.8 & 00 53 55.25 & -72 26 45.8 & 4.4 & B0 & B0.5 [A09] \\
  CH6-20 & O\_6\_311169 & 00 54 09.53 & -72 41 42.9 & 0.62 & 13.71$\pm$0.14 & 0.39$\pm$0.19 & C J005409.57-724143.5 & 00 54 09.57 & -72 41 43.5 & 3.5 & sgB0[e] & sgB0[e] [Z89]\\
  CH6-2 & O\_7\_47103 & 00 54 55.87 & -72 45 10.7 & 0.40 & 15.01$\pm$0.01 & -0.02$\pm$0.01 & C J005455.78-724510.7 & 00 54 55.78 & -72 45 10.7 & 223.4 & B1.5-B3 & B1-1.5 [A09] \\
  CH5-7 & O\_7\_70829 & 00 54 56.17 & -72 26 47.6 & 1.19 & 15.30$\pm$0.01 & -0.04$\pm$0.02 & C J005456.34-722648.4 & 00 54 56.34 & -72 26 48.4 & 23.2 & B0.5 & B0 [A09] \\
  CH5-6 & Z\_2748033 & 00 55 03.63 & -72 22 31.2 & 3.60 & 17.86$\pm$0.03 & -0.03$\pm$0.05 & C J005504.40-722230.4 & 00 55 04.40 & -72 22 30.4 & 38.3 & B1-B5 & unclassified\\
  CH3-18 & Z\_2893439 & 00 56 05.56 & -72 21 59.0 & 0.72 & 15.88$\pm$0.03 & -0.04$\pm$0.03 & C J005605.42-722159.3 & 00 56  05.42 & -72 21 59.3 & 3.1 & B2 & B1 [M08] \\
  CH3-7 & Z\_3075967 & 00 57 24.02 & -72 23 56.4 & 1.30 & 14.71$\pm$0.03 & -0.07$\pm$0.03 & C J005723.77-722357.0 & 00 57 23.77 & -72 23 57.0 & 14.9 & B2 & unclassified \\
  CH3-3 & O\_8\_49531 & 00 57 36.01 & -72 19 33.8 & 0.14 & 16.01$\pm$0.02 & -0.02$\pm$0.04 & C J005736.00-721933.9 & 00 57 36.00 & -72 19 33.9 & 53.7 & B1-B5 & B0-4 [A09]		\\
  XMM1-2 & Z\_4119599 & 01 05 20.72 & -72 49 41.5 & 4.01 & 16.98$\pm$0.03 & -0.09$\pm$0.04 & X J010519.9-724943 & 01 05 19.90 & -72 49 43.1 & 14.6 & B3-B5 & unclassified\\
  XMM1-3 & Z\_4232476 & 01 06 21.02 & -72 40 48.8 & 4.53 & 16.38$\pm$0.03 & 0.02$\pm$0.03 & X J010620.0-724049 & 01 06 20.01 & -72 40 49.1 & 35.9 & B9 & unclassified\\
  \hline
   \label{t-source_list}

   \end{tabular}
  \end{minipage} 

 \begin{flushleft}

   (1) Source ID - the name convention is CH/XMM F-NN, where CH stands for \textit{Chandra} and XMM for \textit{XMM-Newton} sources, F is the field number and NN is the source ID for this field (from \citealt{Antoniou09bb} and \citealt{Antoniou10}, respectively). \\
   (2) Optical counterpart ID given as O\_F\_NNNNNN from the OGLE-II catalog \citep{Udalski98}, where F is the field number and NNNNNN is the optical source ID respectively, and Z\_NNNNNN from the MCPS catalog \citep{Zaritsky02}, where NNNNNN is the line number of the source in their table 1.\\
   (3,4) Right Ascension and Declination (J2000) of the optical counterpart.\\
   (5) Distance (in arcseconds) of the optical counterpart from the X-ray source. \\
   (6,7) Apparent \textit{V} magnitude and \textit{B-V} color index, along with their errors, taken directly from the original catalogs without applying any reddening or zero-point correction. \\
   (8) X-ray source names, \textit{Chandra} sources (CXOU) labeled as C and \textit{XMM-Newton} sources (XMMU) as X, followed by the designation.\\
   (9,10) Right Ascension and Declination (J2000) of the X-ray source.\\
   (11) Unabsorbed X-ray luminosity in the 0.5-7.0 keV band (assuming a power law with photon index $\Gamma = 1.7$ and $\rm {N_H=6\times10^{20}}$ cm$^{-2}$, and 60 kpc distance; \citealt{Hilditch05}) for \textit{Chandra} sources (Zezas, in prep.; \citealt{Antoniou09bb}) and \textit{XMM-Newton} sources \citep{Antoniou10}. \\
   (12) Spectral-type classification from this work. \\
   (13) Classification published in other works: [A09]: \citet{Antoniou09}; [M08]: \citet{McBride08}; [Z89]: \citet{Zickgraf89}. \\
   $*$: the classification of source CH7-19 is tentative as its $<$FWHM$_{\rm {H}_\alpha}>$ is marginally lower than the limit set for BeXRBs.
   \\

 \end{flushleft} 
 \end{table}
\end{landscape} 
}

\begin{table}
 \caption{Measured H$\alpha$ equivalent widths (\textit{EW}) for each source on July 26, 2008$^{(1)}$.}
 \begin{tabular}{ l c c c }
 \hline\hline
 Source ID & $EW_{H\alpha}$ & $\pm \delta(EW_{H\alpha})$ \\
           &  ({\AA}) & ({\AA}) \\ 
 \hline
 CH4-8 & -6.45 & 0.07 \\
 CH7-1 & -13.78 & 0.13 \\
 CH4-2 & -24.22 & 0.52 \\ 
 CH4-5 & -29.39 & 0.17 \\
 CH7-19 & -187.68$^{(2)}$ & 0.36 \\
 CH4-3 & -23.19 & 0.08 \\
 CH5-3 & -2.90 & 0.04 \\
 CH5-1 & -8.54 & 0.06 \\
 CH6-1 & -7.65 & 0.10 \\
 CH5-12 & -4.41 & 0.06 \\
 XMM2-1 & -21.06 & 0.09 \\
 CH5-16 & -12.31 & 0.04 \\
 CH6-20 & -120.3 & 1.6 \\
 CH6-2 & -42.68 & 0.14 \\
 CH5-7 & -14.89 & 0.13 \\
 CH5-6 & -12.31 & 0.04 \\
 CH3-18 & -32.87 & 0.10 \\
 CH3-7 & -1.01 & 0.05 \\
 CH3-3 & -33.35 & 0.18 \\
 XMM1-2 & -53.63 & 0.23 \\
 XMM1-3 & -6.59 & 0.14 \\
 \hline  
 \end{tabular}
 
 $^{(1)}$ For source CH4-2 the \textit{EW} was measured on September 19, 2008. \\ 
 $^{(2)}$ Part of the H$\alpha$ emission line could be due to ISM contamination.
 \label{t-ha_ew_list}
\end{table}

\begin{figure}
\resizebox{\hsize}{!}{\includegraphics{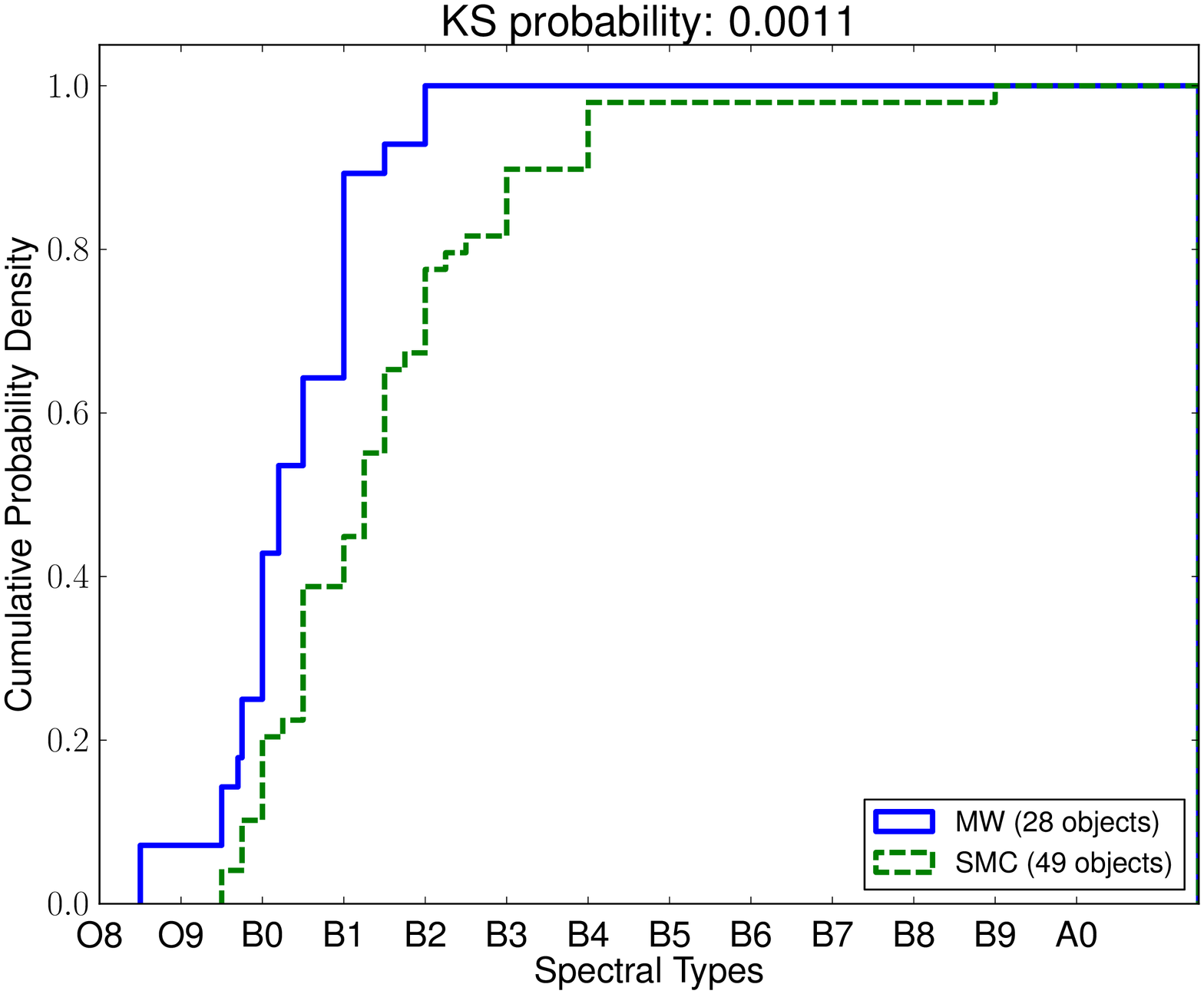}}
\resizebox{\hsize}{!}{\includegraphics{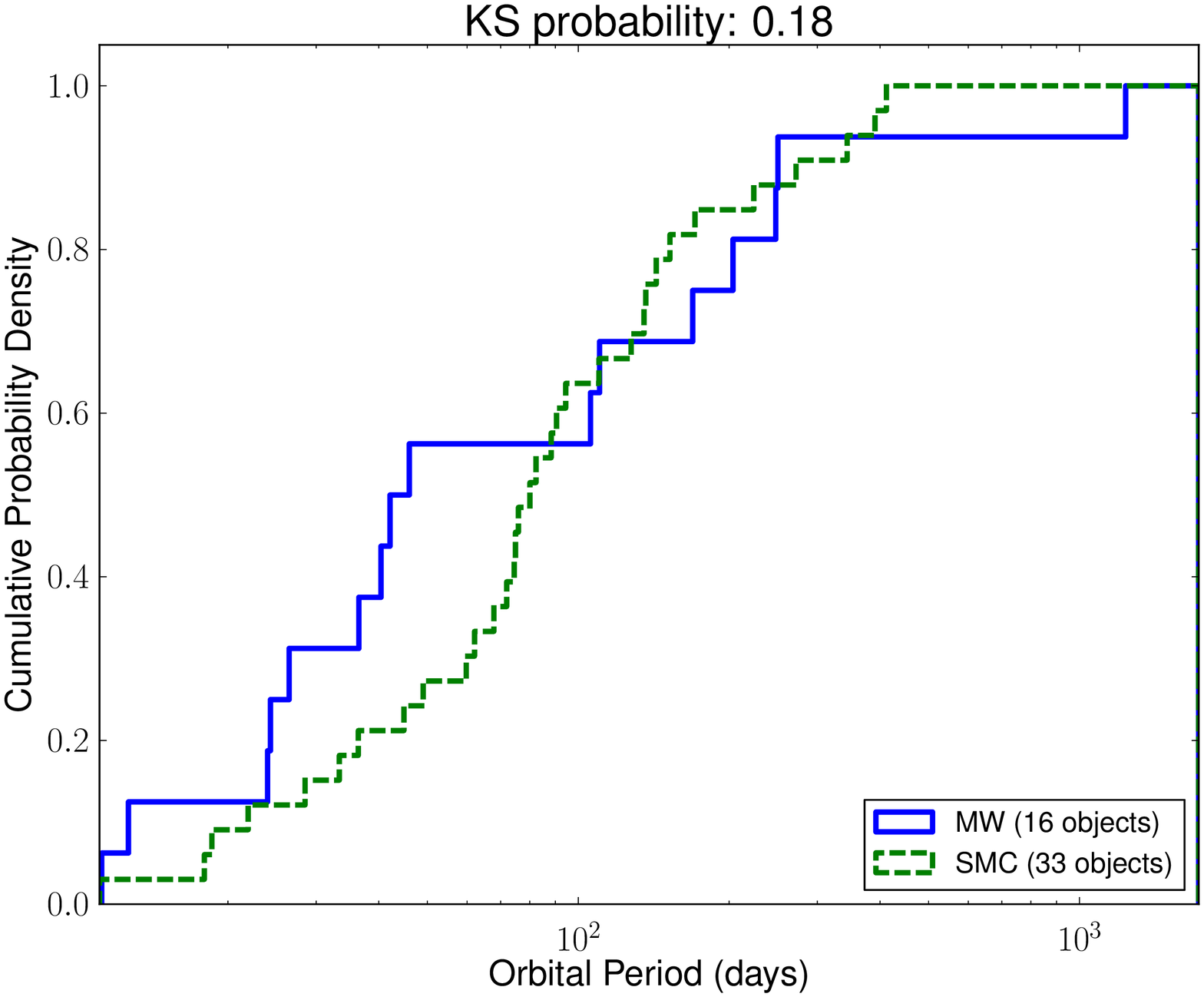}}
\resizebox{\hsize}{!}{\includegraphics{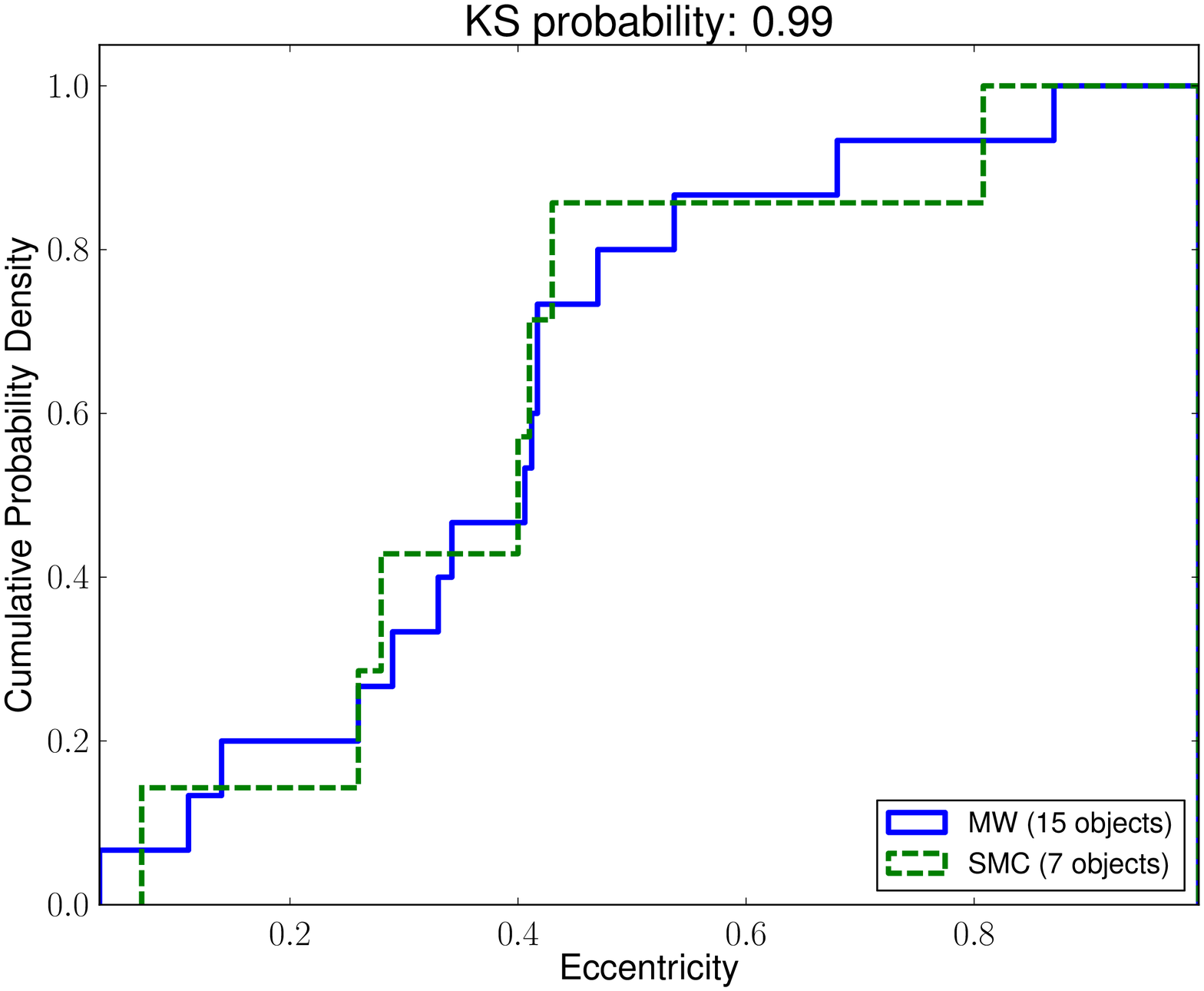}}
\caption{Cumulative distributions of the spectral types, orbital periods ($P_{orb}$), and eccentricities (\textit{e}) for BeXRBs in the MW (solid blue) and the SMC (dashed green).    
For the spectral-type distributions the data come from this work, \citet{McBride08}, and \citet{Antoniou09},  while for the MW we take the data from the review paper of \citet{Reig11}. Data for the $P_{orb}$ of BeXRBs in the SMC are taken from \citet{Rajoelimanana11}, and for the MW from \citet{Townsend11}. For the \textit{e} we take all data (for both SMC and MW) from \citet{Townsend11}. The Kolmogorov-Smirnov test probabilities are given over each plot. We find marginal evidence for difference between the spectral type distributions of SMC and MW BeXRBs (at $\sim$99.9\% confidence), while we do not find any evidence for difference in the distribution of the orbital parameters ($Porb$, $e$) in SMC and MW BeXRBs (see Section \ref{s-spectral_distributions}).   
}
\label{f-kstests-orbital_param}
\end{figure}

In Fig. \ref{f-kstests-orbital_param} we present the cumulative distributions of the spectral types, orbital periods, and eccentricities for MW and SMC BeXRBs. For the SMC we used the spectral types of sources as derived from this work and the previous studies of \citet{McBride08}, and \citet{Antoniou09}, while for the MW we take the data from the review paper of \citet{Reig11}. We find marginal evidence for difference between the spectral type distributions of SMC and MW BeXRBs ($\sim$99.9\% confidence based on the KS test), in contrast to previous studies which found no evidence for difference between the BeXRB populations in these two galaxies (e.g. \citealt{McBride08,Antoniou09}). We attribute this difference to the use of larger samples for both the SMC and the MW than previous studies, and the extension of the former to later spectral types. However, given the important implications of this difference for the evolution of BeXRBs in the SMC we consider it as only tentative at this point. 
 On the other hand, we do not find any evidence for difference in the distribution of the orbital parameters ($Porb$, $e$) in SMC and MW BeXRBs, based on the KS test. This result provides an indication for the distribution of the kick velocities imparted on the pulsars during the supernova explosion. The vector of the kick and the mass of the secondary object (in the initial binary) will affect the orbit of the neutron star around it. Since the spectral types (which reflects the donor mass of the BeXRBs) are not dramatically different between the SMC and the MW, and the orbital-element distributions (period and eccentricity) are not different at a statistically significant level, we can deduce that there is an indication that the kicks imparted to the pulsars during the supernova explosions have similar strengths in the SMC and the MW. Further support comes from previous studies, where the measured kicks in the MW BeXRBs ($v\sim15\pm6$ km s$^{-1}$; \citealt{vandenHeuvel00}) are found to be comparable with these of the SMC BeXRBs ($v\sim30$ km s$^{-1}$; \citealt{Coe05b}, $v<15-20$ km s$^{-1}$; \citealt{Antoniou10}).

\subsection{The case of the supergiant B[e] source CXOU J005409.57-724143.5}
\label{s-sgcase}

\subsubsection{Optical and Infrared properties}
\label{s-sgcase_optirprop}

\begin{figure}
\centering
\resizebox{\hsize}{!}{\includegraphics{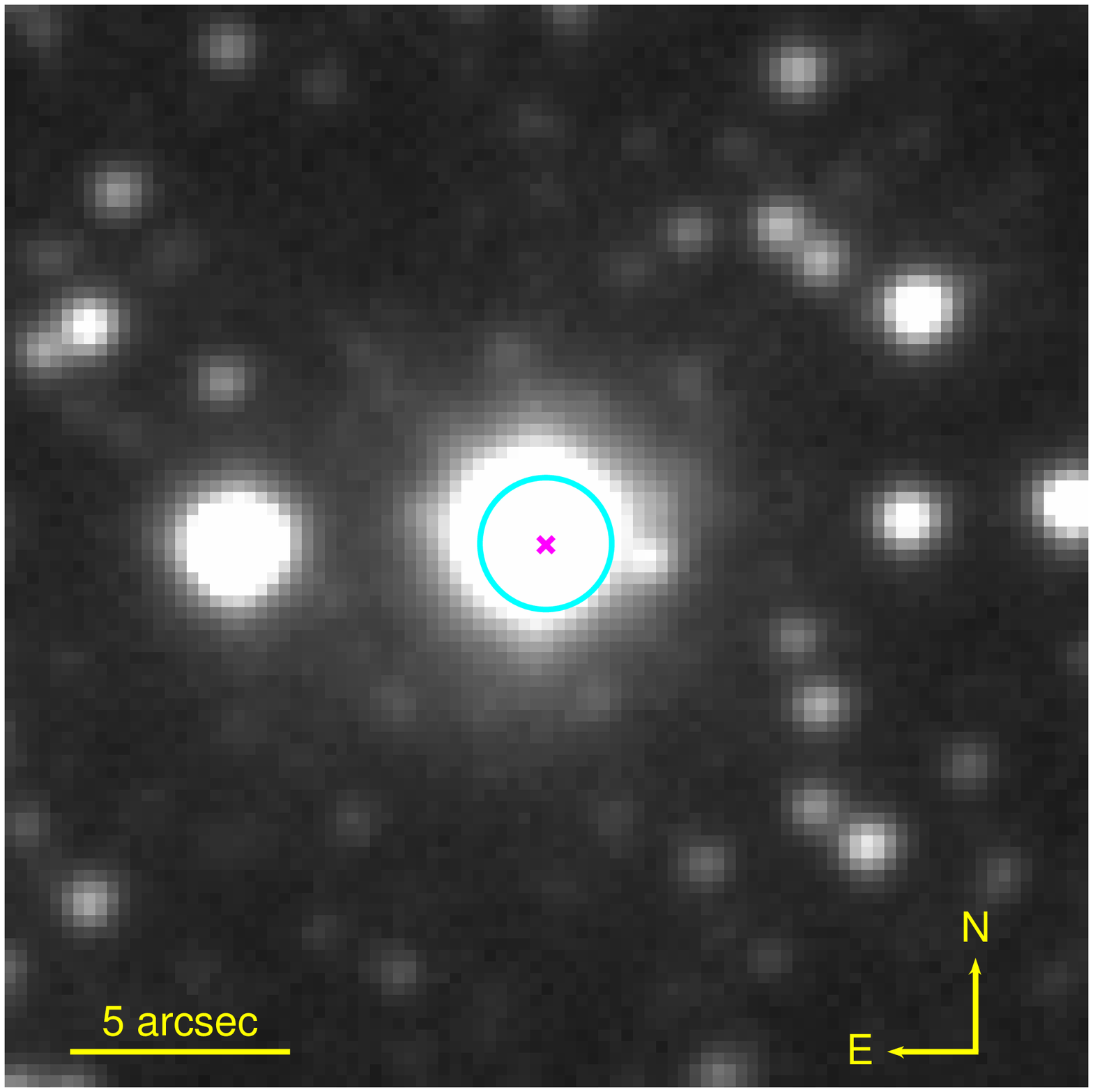}}
\caption{Finding chart of source CXOU J005409.57-724143.5 (or CH6-20 in \citealt{Antoniou09bb}) from an OGLE-III $I$-band image \citep{Udalski08}. The dimensions of the field are 24.7\arcsec$\times$24.7\arcsec. The cyan circle indicates the positional error-circle of the \textit{Chandra} source (1.5\arcsec radius, see \citealt{Antoniou09bb} for details) and the magenta x-symbol indicates the location of star number 7 from the OGLE-III SMC105.6 map, which is spatially coincident with star LHA 115-S 18 \citep{Henize56}.
}
\label{f-finding_chart}
\end{figure}

The optical counterpart of source CXOU J005409.57-724143.5 (CH6-20) is classified in this work as a supergiant star. If the X-ray source is an accreting binary, it would be only the second to be found in the SMC after SMC X-1 \citep{Webster72}. Its optical counterpart is star LHA 115-S 18 \citep{Henize56} also known as AzV 154 \citep{Azzopardi75}, a well known bright emission-line star that has been studied systematically since 1956 (\citealt{Shore87,Zickgraf89,Morris96,Massey01,vanGenderen02}, and references therein), and more recently in \citet{Clark13}. In Fig. \ref{f-finding_chart} we present a finding chart of the X-ray source and its corresponding optical counterpart. 

According to the classification criteria for supergiant B[e] stars \citep{Lamers98}, these objects exhibit strong Balmer emission lines (usually with P Cygni profiles indicating mass loss), low excitation permitted and forbidden lines (e.g. FeII), and strong near or mid-infrared excess (due to hot circumstellar dust). Moreover, they are rather stable photometrically (variation of the order of $\sim$0.1-0.2 mag) and spectroscopically, unlike the S Doradus / Luminous Blue Variable (LBV) stars which exhibit similar spectra \citep{Zickgraf86}. 

As discussed in Section \ref{s-discus_sources}, star S 18 presents the typical spectral characteristics of B[e] stars. In addition, it is very bright in the near-IR showing a very large color excess of $J$-[3.6$\mu$m]=$3.17\pm0.05\rm \,mag$, compared to the main sequence and supergiant B stars (most Be stars show a color excess of $J$-[3.6$\mu$m]$\sim0.8\rm \,mag$; \citealt{Bonanos10}). 

\begin{figure}
\resizebox{\hsize}{!}{\includegraphics{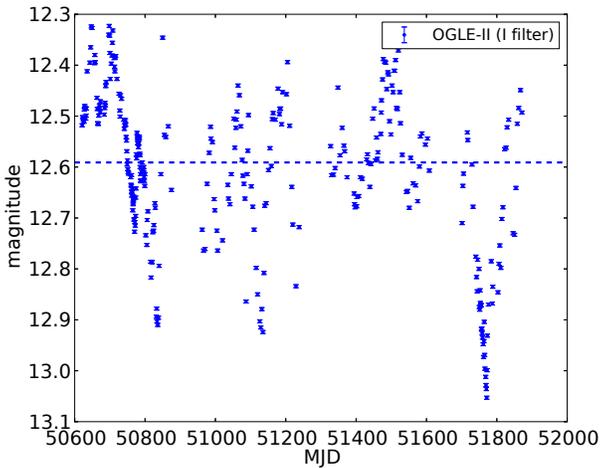}}
\caption{ 
OGLE-II $I$-band light curve of the optical counterpart of source CH6-20 (\textit{ROSAT} observations were obtained before the OGLE-II survey, while \textit{Chandra} and \textit{XMM-Newton} observations were obtained later). The horizontal line indicates the median value of $\sim12.6$ mag, and the errors on the photometric points are 0.003 mag. The observed aperiodic variability is larger than what expected for a normal supergiant B[e] star and it is more similar to the variability that Luminous Blue Variables display (see Section \ref{s-sgcase_optirprop}).
}
\label{f-sg_lc}
\end{figure}

In Fig. \ref{f-sg_lc} we present the light curve of star S 18, using the OGLE-II data in the $I$ filter (as discussed in Section \ref{s-optir_data}), clearly indicating a highly variable source. In addition, S 18 exhibits spectroscopic variations (e.g. in the lines of HeII $\lambda4686$, CIV $\lambda1550$, NIV $\lambda1487$; \citealt{Shore87}). This is in contrast with the photometric stability of typical sgB[e] stars, and it is more similar to the strong variability of LBV stars (e.g. \citealt{Zickgraf86}). However, there is growing evidence that the sgB[e] and the LBV sources are not distinct classes \citep{Morris96,vanGenderen02,Clark13}. In order to investigate if it is an accreting binary source we searched for modulation of its optical emission resulting from an orbital period. Due to the significant variability, we detrended the light curve following \citet{Schurch11}, by fitting a 2nd order polynomial to each year-long segment of the data. Then we obtained the Lomb-Scargle periodogram \citep{Lomb76,Scargle82} for the entire detrended light curve. We did not find any periodicity at the 90\% confidence level in agreement with the results of \citet{Clark13}. The significance level was estimated by simulating light curves based on the noise characteristics of the data and repeating the analysis for each simulated light curve.

\begin{figure}
\resizebox{\hsize}{!}{\includegraphics{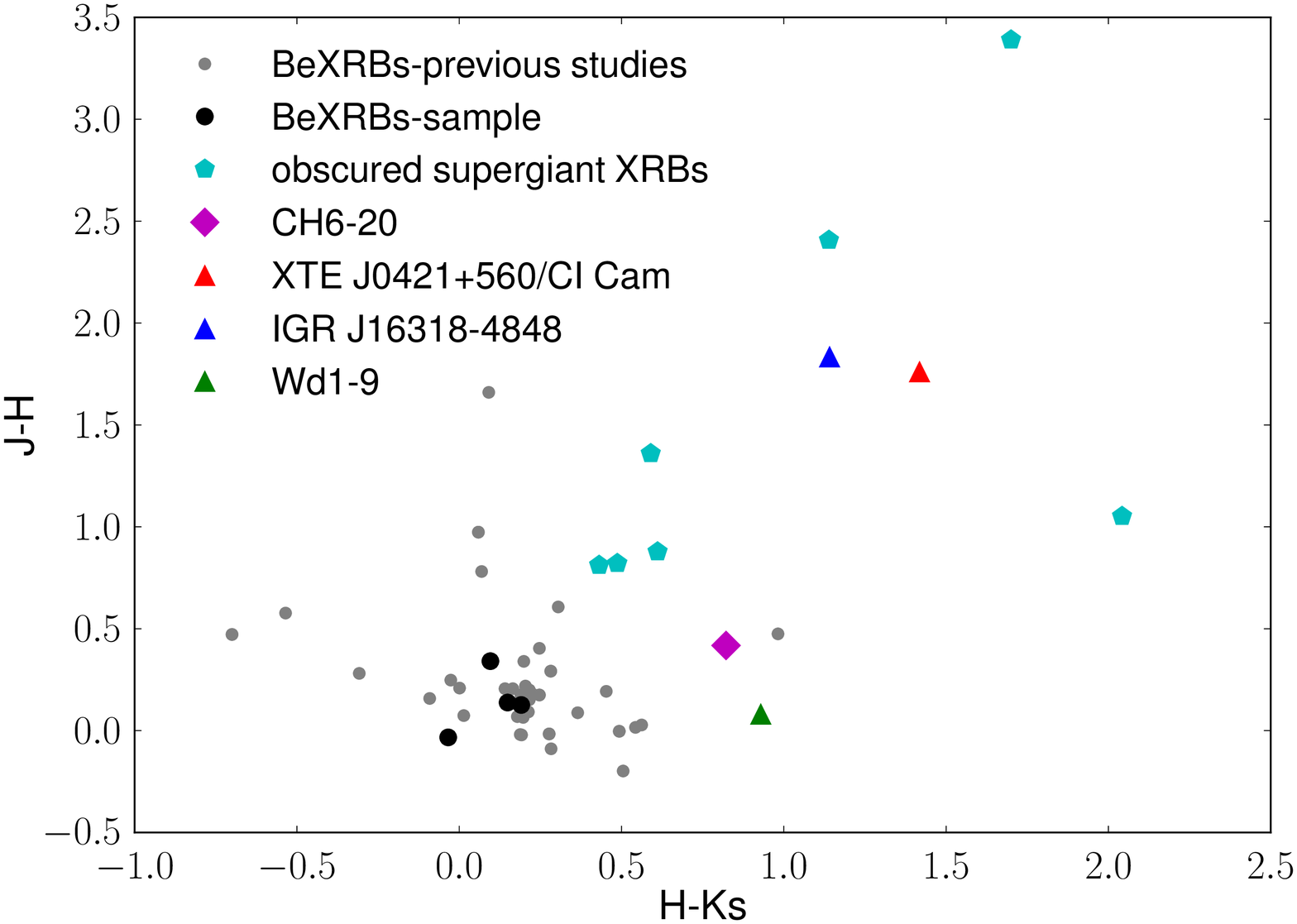}}
\caption{$J$-$H$ vs. $H$-$K_{s}$ color-color diagram of BeXRBs with available 2MASS data. Black dots show sources from our sample and gray dots show sources from the samples of \citet{McBride08}, and \citet{Antoniou09}. We present also X-ray sources with a known sgB[e] companion (triangles) and the highly obscured sgXRBs identified so far by \textit{INTEGRAL} (cyan polygons; \citealt{Manousakis11b}). Our source CH6-20 is presented as a purple diamond.
}
\label{f-ir_excess}
\end{figure}

\subsubsection{X-ray properties}
\label{s-sgcase_xrayprop}

In order to explain the spectral variations of S 18, \citet{Shore87} suggested the presence of a hot companion, possibly a helium star or a neutron star. But no further evidence of such an object existed as the X-ray observations at the time were not sensitive enough to observe any emission from accretion onto a putative compact object. Nowadays, \textit{Chandra} and \textit{XMM-Newton} are able to routinely detect sources with X-ray luminosities down to $\sim10^{33}$ erg s$^{-1}$ at the distance of the SMC (with 10-30 ks exposures) and have succeeded in detecting source CXOU J005409.57-724143.5 (CH6-20) which is spatially coincident with star S 18. In the study of \citet{Antoniou09bb} its proposed optical counterpart is star O\_6\_311169 (00:54:09.53, -72:41:42.9) from the OGLE-II catalog \citep{Udalski98}, or Z\_2611188 (00:54:09.57, -72:41:42.9) from the Magellanic Clouds Photometric Survey (MCPS) catalog \citep{Zaritsky02}, which is coincident with star S 18. With this work we confirm that the optical counterpart of the source CXOU J005409.57-724143.5 (CH6-20) is the supergiant B[e] star S 18. 

This source has only been detected in the X-ray band in only two epochs (see Section \ref{s-xray_data}). The lack of previous detections with \textit{ROSAT} or \textit{Einstein} indicate that it is persistently at a low luminosity state (${\rm L_x}<10^{35}$ erg s$^{-1}$) since at higher luminosities it would have been detected already in one of the several pointings with these observatories. 
Although the intensity of CH6-20 is only $\sim2\sigma$ above the background in the \textit{Chandra} survey, the independent \textit{XMM-Newton} detection with a source-detection likelihood of DET\_ML=24.6 during the first \textit{XMM-Newton} observation ensures that the source is not spurious. 

Next we discuss the X-ray properties of this source in the context of typical supergiant wind-fed X-ray binaries. Assuming Bondi-Hoyle accretion from a stellar wind, the X-ray luminosity of the compact object is given by (e.g. \citealt{Longair11}):
\begin{eqnarray}
 {\rm L_X} \sim \eta \frac{\dot{M}c^2}{4}\left( \frac{2GM_X}{\alpha} \right)^2 v_w^{-4}, 
\label{eq-lx}
\end{eqnarray} 
where $\eta$ is the efficiency of the conversion of the rest mass energy of the accreted matter into radiation ($\sim0.1$), 
$\dot{M}$ is the mass-loss rate, $c$ is the speed of light, $G$ is the gravitational constant, $M_X$ is the mass of the compact object, $\alpha$ is the separation between the compact object and the donor star, and $v_w$ is the wind velocity. The distance $\alpha$ can be calculated at the periastron from Kepler's 3rd law:
\begin{eqnarray}
 \alpha = \left( {\frac{G}{4\pi^2} \left( M_X + M_{OB} \right) P_{orb}^2} \right) ^{\frac{1}{3}} \times (1-e), 
\end{eqnarray}
where $M_{OB}$ is the mass of the donor OB star, $P_{orb}$ is the orbital period of the system, and \textit{e} is the eccentricity. For the donor star, \citet{Zickgraf89} estimate a mass of $M_{OB}\sim 40$ M$_{\odot}$, a wind mass-loss rate of $\dot{M}\sim3 \times 10^{-5}$ M$_\odot$ yr$^{-1}$, and a velocity of $v_w\sim750$ km s$^{-1}$. Since we have no information on the orbital parameters of the system we take the typical (median) values for the orbital period ($P_{orb} \sim 6.78$ d) and eccentricity ($e \sim$ 0.17) for other known sgXRBs (from the data in table 7 of \citealt{Townsend11}, for SMC and Milky Way systems). This results in a peak X-ray luminosity at periastron of $\rm L_X \sim 10^{37}$ erg s$^{-1}$. Keeping the same values for the orbital period and eccentricity, but using instead the typical wind parameters for supergiant stars in the SMC within the spectral range O9.5-B1 ($\dot{M}\sim 8.4 \times 10^{-7}$ M$_\odot$ yr$^{-1}$ and  $v_w\sim1272$ km s$^{-1}$; \citealt{Mokiem07}) we derive a luminosity of $\rm L_X \sim 7 \times 10^{34}$ erg s$^{-1}$.

The estimated peak $\rm L_X$ at periastron from the values obtained from \citet{Zickgraf89} is too high for this source to remain undetected for so long. On the other hand, when the values from \citet{Mokiem07} are used, the estimated peak $\rm L_X$ is consistent with the measured luminosity and comparable to the detection limits of the \textit{Chandra} and \textit{XMM-Newton} observations. However, the above luminosities correspond to the peak $\rm L_X$ for this source when the neutron star is located at periastron, assuming a stable mass-loss rate. If the X-ray observations were obtained when the neutron star is away from the periastron and/or a lower mass-loss rate phase (as for example it would be expected from its significant optical variability)
then the X-ray luminosity would be even lower than that estimated above. 

Furthermore, a clumpy wind with a high mass-loss rate may provide an alternative interpretation for the nature of this source. In this case the high column density towards the neutron star due to the wind may obscure its X-ray emission. A column density of $\rm {N_H}=3\times10^{24}$ cm$^{-2}$ (assuming a power-law spectrum with a photon index $\Gamma$=1.7 in the 0.5-7 kev energy band) would be enough to attenuate a source with intrinsic $\rm L_X=10^{37}$ erg s$^{-1}$ (which is the upper range in its X-ray luminosity based on the parameters from \citealt{Zickgraf89}) down to $\rm L_X=5.7\times10^{33}$ erg s$^{-1}$, while an $\rm {N_H}=1\times10^{24}$ cm$^{-2}$ would be enough to attenuate a source with intrinsic $\rm L_X=7\times10^{34}$ erg s$^{-1}$ (the luminosity expected from the parameters of \citealt{Mokiem07}) down to $\rm L_X=1.6\times10^{33}$ erg s$^{-1}$. Both of these estimates are half the value we found with \textit{Chandra}. However, we do have significant detection below 2 keV in both the \textit{Chandra} and \textit{XMM-Newton} observations ($6_{-2}^{+3}$ 
and 39$\pm$11 counts respectively), and only upper limits or marginal detections above 4 keV,
which do not allow us to set any useful constraints on the spectral parameters of this source.

There is an emerging subclass of highly obscured wind-fed sgXRBs, discovered in recent \textit{INTEGRAL} observations \citep{Walter06}. To date there are three known X-ray sources with B[e] companions: IGR J16318-4848 \citep{Walter03,Filliatre04}, which is considered the prototype of highly obscured wind-fed sgXRBs, CI Cam/XTE J0421+560 \citep{Clark99,Boirin02}, the first HMXB with a sgB[e] companion, and Wd1-9 \citep{Clark08}, a probable colliding-wind system. Recently, \citet{Bartlett12} presented a detailed study of CI Cam observed in 2003 with \textit{XMM-Newton}. This system has a B0-2[e] supergiant companion  \citep{Clark99} and shows a heavily absorbed ($\rm {N_H} \sim 4.4\times10^{23}$ cm$^{-2}$) power-law ($\Gamma \sim 1$) spectrum reaching $\rm L_X\sim4.1\times10^{33}$ erg s$^{-1}$ (3-10 keV) in quiescence. 
In 1998 it showed an outburst reaching a flux of $\sim 4.8 \times 10^{-8}$ erg cm$^{-2}$ s$^{-1}$ (2 Crab) in the 2-10 keV energy band \citep{Smith98}. Assuming a distance of 5 kpc, the outburst luminosity is $\sim 5.7 \times 10^{36}$ erg s$^{-1}$. The nature of the compact object in this system as well as in IGR J16318-4848 is unclear \citep{Bartlett12}, although there is some indication that at least in the case of IGR J16318-4848 it might be a neutron star \citep{Filliatre04}.

 The main difference of these systems from the "classical" sgXRBs is that they are much more absorbed in the X-ray band. The compact object is deeply embedded in the dense material of the wind and the absorbing column density changes during the orbit (it may increase up to ten times close to the eclipse; \citealt{Manousakis11}). Moreover, material away from the compact object (like the material around the sgB[e] star) may contribute to the absorption (e.g. as in IGR J16318-4848; \citealt{Walter06}). In this case the source exhibits strong IR excess, attributed to free-free and bound-free emission from hydrogen in a circumstellar envelope (or disk; \citealt{Wisniewski07}). At longer wavelengths such as those observed by \textit{Spitzer}, the IR excess indicates the presence of warm dust in the circumstellar envelope or the disk. In Fig. \ref{f-ir_excess} we present the location of source CH6-20 in a near infra-red (NIR) color-color diagram. This plot is an updated version of fig. 4 by \citet{Graus12}, where in addition to the BeXRBs we include all highly obscured sgXRBs identified so far by \textit{INTEGRAL}. These sources show luminosities in the $10^{33}-10^{34}$ erg s$^{-1}$ range, they are highly variable and they have absorbing column densities in the $10^{23}-10^{24}$ cm$^{-2}$ range. We find that source CH6-20 shows redder NIR colors than the other BeXRBs which place it closer to the colors of the obscured supergiant HMXBs. 

An alternative scenario for the nature of the CH6-20 is that it is a colliding-wind binary (e.g. \citealt{Pollock87}). These systems consist of two orbiting massive stars in orbit, and their low luminosity X-ray emission ($\rm L_X\sim10^{33}-10^{34}$ erg s$^{-1}$) is produced by shocks at the collision front of the winds. \citet{Clark13} point out that although S 18 (CH6-20) and the wind-fed sgXRB CI Cam exhibit many similarities, their long term photometric and spectroscopic behaviors are not quite the same, and argue for a colliding-wind nature for the former system instead of an accreting binary. This is supported by the location of source CH6-20 in the NIR color-color diagram (Fig. \ref{f-ir_excess}) which is also consistent with that of the colliding-wind system Wd1-9 \citep{Clark08} and their similar X-ray luminosity (unabsorbed $\rm L_X\sim4\times10^{33}$ erg s$^{-1}$ for Wd1-9, in the 0.5-8 keV band; \citealt{Clark08}). The luminosities of both CH6-20 and Wd1-9 are well within the range of colliding-winds systems ($\sim10^{32}-10^{35}$erg s$^{-1}$, \citealt{Gagne12,Stevens92}). Although colliding-wind systems also have hard X-ray spectra, a possible discriminating feature from accreting pulsar binaries is that the spectra of the former have a cutoff at $\sim$10 keV, and they show effectively no photons with energies above 10 keV (Pollock, priv. comm.). Unfortunately the weak X-ray emission of this source does not allow us to detect any X-ray photons above 2 keV. This in combination with its highly variable optical light curve that does not allow us to measure its orbital parameters, hamper the distinction between these two scenarios. However, if the heavily obscured sgB[e] scenario proves to be correct then CXOU J005409.57-724143.5 (CH6-20) / S 18 will be the first extragalactic heavily obscured sgXRB, and the second sgXRB in the SMC.

\section{Conclusions}
\label{s-conclusions}

In this paper we presented our results from optical spectroscopic observations of the optical counterparts of X-ray sources detected in the \textit{Chandra} and \textit{XMM-Newton} surveys of the SMC. We used the AAOmega spectrograph at the Anglo-Australian Telescope to observe sources identified in Zezas (in prep.) and \citet{Antoniou09bb,Antoniou10}, with the aim to identify new HMXBs and determine their spectral types. We identified 5 new BeXRBs and 1 known supergiant system which we associate with an X-ray source. We confirmed the previous classifications (within 0.5 spectral type) of 12 sources, while for 3 sources our revised classification, with higher resolution and S/N data, result in later (by 1-1.5 subclass) spectral types.

We were able to classify BeXRBs with X-ray luminosities over 3 orders of magnitude. The selection of the parent sample from \textit{Chandra} and \textit{XMM-Newton} observations of the SMC, allows us to extend our census of BeXRBs to almost quiescent luminosities. A comparison of the populations of BeXRBs in the SMC and the Milky Way with respect to their spectral types reveals a marginal evidence for difference. However, we find no statistically significant differences for their orbital parameters (periods and eccentricities). This result further supports other lines of evidence for similar supernova kick velocities between the low metallicity SMC and the Milky Way.  

Finally, we discuss the X-ray, optical, and infrared properties of source CXOU J005409.57-724143.5. This intriguing source is associated with the well known supergiant star LHA 115-S 18. Its optical and X-ray properties do not allow us to distinguish between a colliding-wind system or a supergiant X-ray binary. If the second scenario proves to be correct, then this source would be the first extragalactic supergiant X-ray binary with a B[e] companion.

\subsubsection*{Acknowledgments}    
We would like to thank the referee, Phil Charles, whose comments and suggestions helped to improve the paper. We would like to thank the staff of the Anglo-Australian Observatory for obtaining the data used in this work. We thank A. Manousakis and R. Walter for useful discussions regarding the class of heavily obscured sgXRBs, and A. Pollock for advice regarding the class of colliding-wind binaries. GM  acknowledges support by State Scholarships Foundation of Greece in the form of a scholarship. AZ acknowledges support by the \textit{Chandra}  grant GO3-14051X, the NASA grant NNX12AL39G, and the EU IRG grant 224878. Space Astrophysics at the University of Crete is supported by EU FP7-REGPOT grant 206469 (ASTROSPACE). VA acknowledges support by NASA grant NNX10AH47G issued through the Astrophysics Data Analysis Program, and \textit{Chandra}  grant GO3-14051X.

This research has made use of NASA's Astrophysics Data System, SIMBAD database (operated at CDS, Strasbourg, France), and data products from the Optical Gravitational Lensing Experiment and the Two Micron All Sky Survey, which is a joint project of the University of Massachusetts and the Infrared Processing and Analysis Center/California Institute of Technology, funded by the National Aeronautics and Space Administration and the National Science Foundation. Figures were generated using the matplotlib\footnote{\url{http://matplotlib.org/}} library in the Python\footnote{\url{http://python.org/}} programming language.

\appendix
\section{Supplementary spectral classification results}
\label{s-suppl_spectral_results-appendix}

In this section we review the spectral classification for the sources not discussed in Section \ref{s-discus_sources} (15 in total). For sources that we confirm previous classifications we only give the spectral classification, while for sources with updated classifications we give a detailed account of the basis for these new classifications. We also present all spectra in Fig. \ref{f-extra_spectra}.

\begin{figure*}
\includegraphics[scale=0.59]{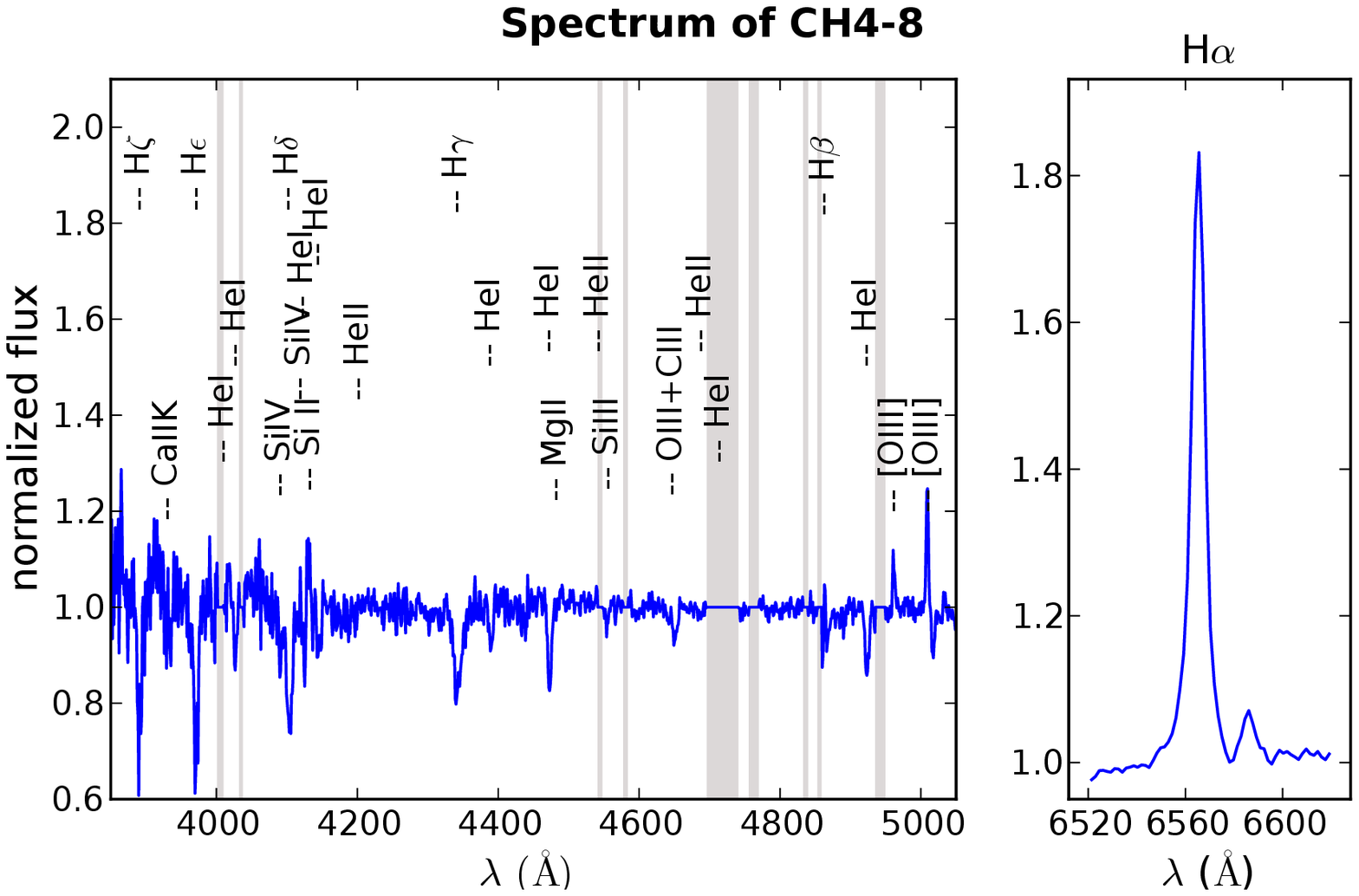} \\
\includegraphics[scale=0.59]{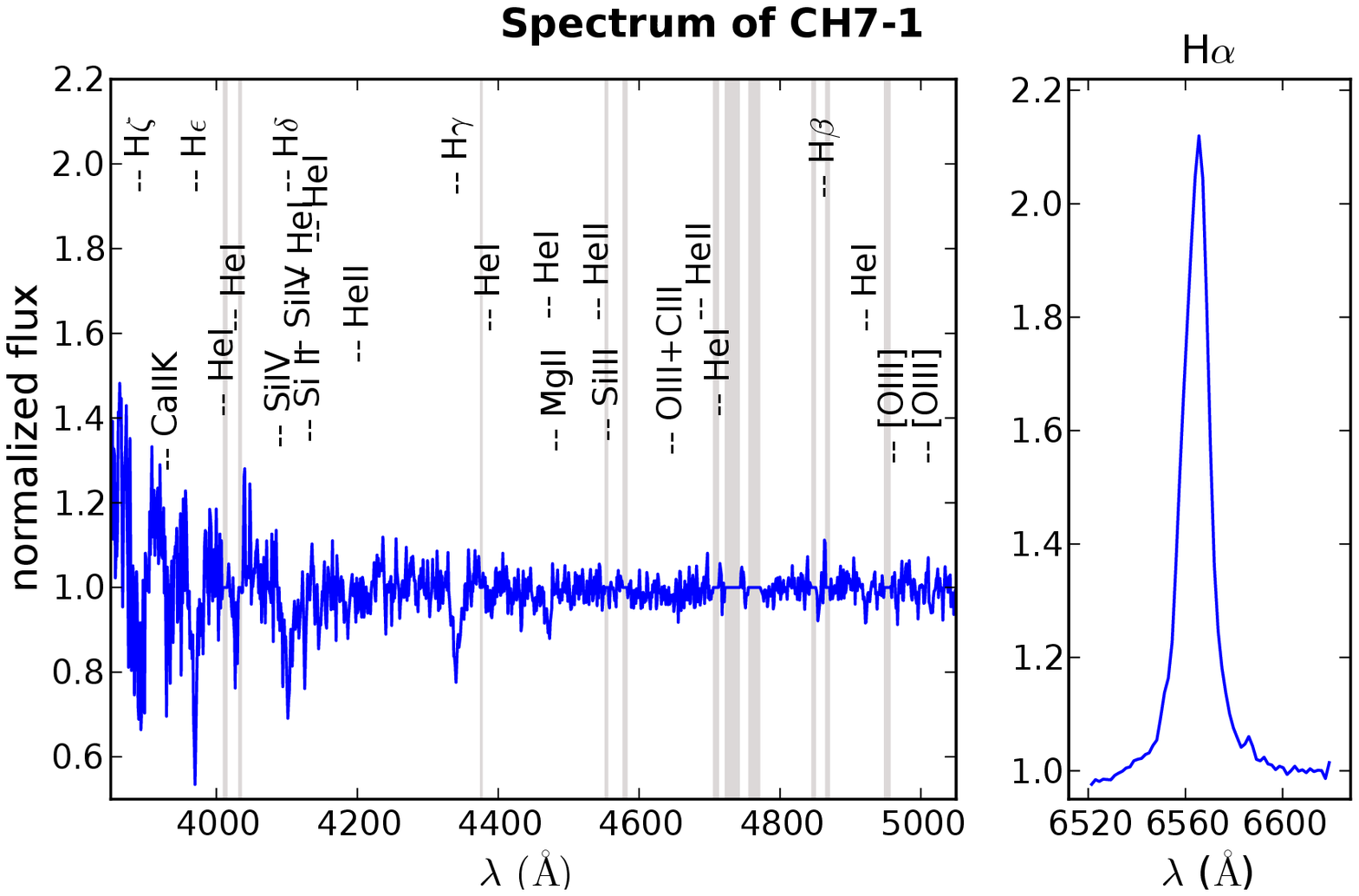} \\
\includegraphics[scale=0.59]{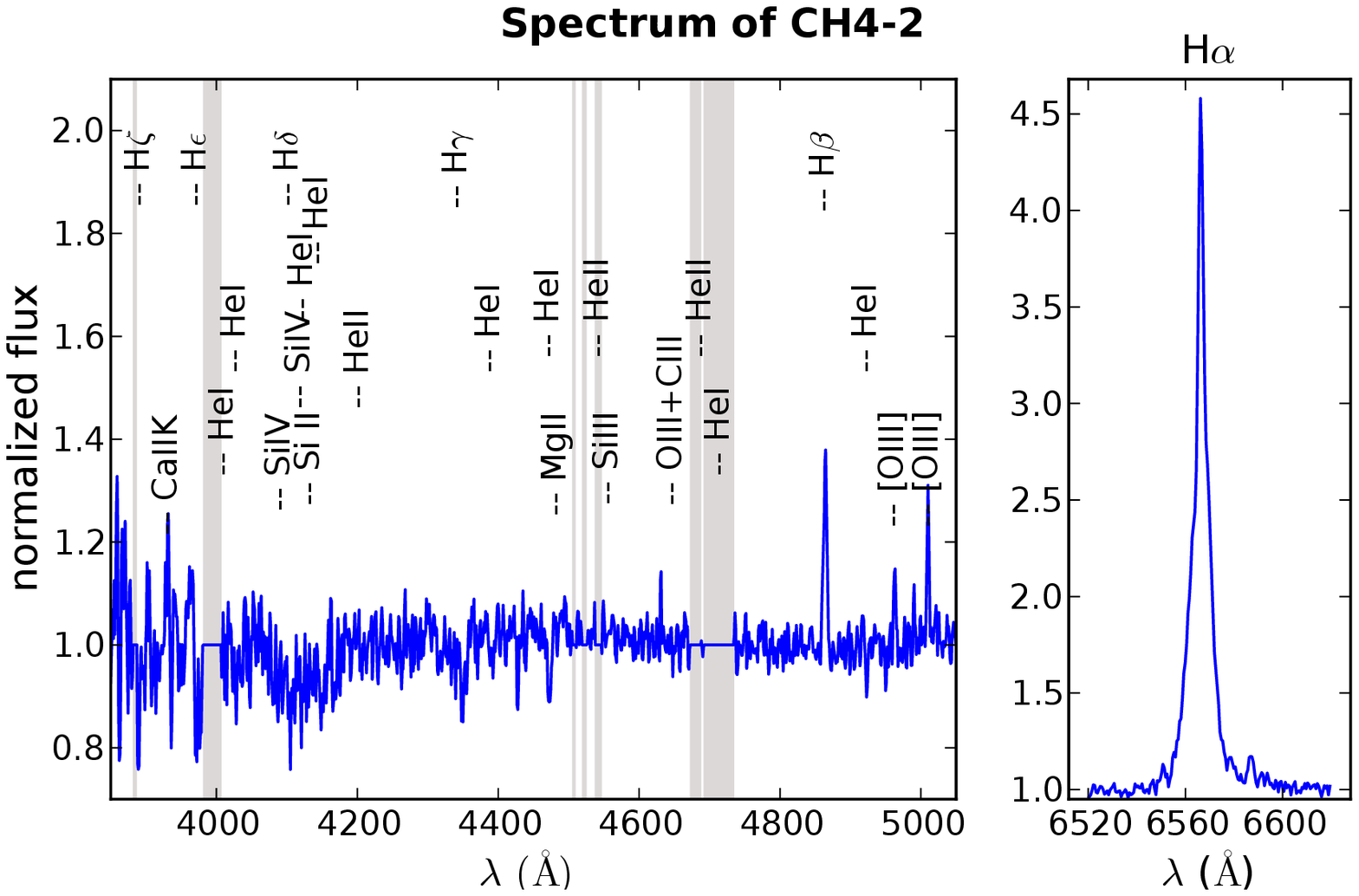} \\
\caption{The spectra of BeXRBs studied in this work with previously known classifications. Shaded areas indicate wavelength ranges for bad columns and/or sky subtraction residuals.
}
\label{f-extra_spectra}
\end{figure*}

\begin{figure*}
\includegraphics[scale=0.59]{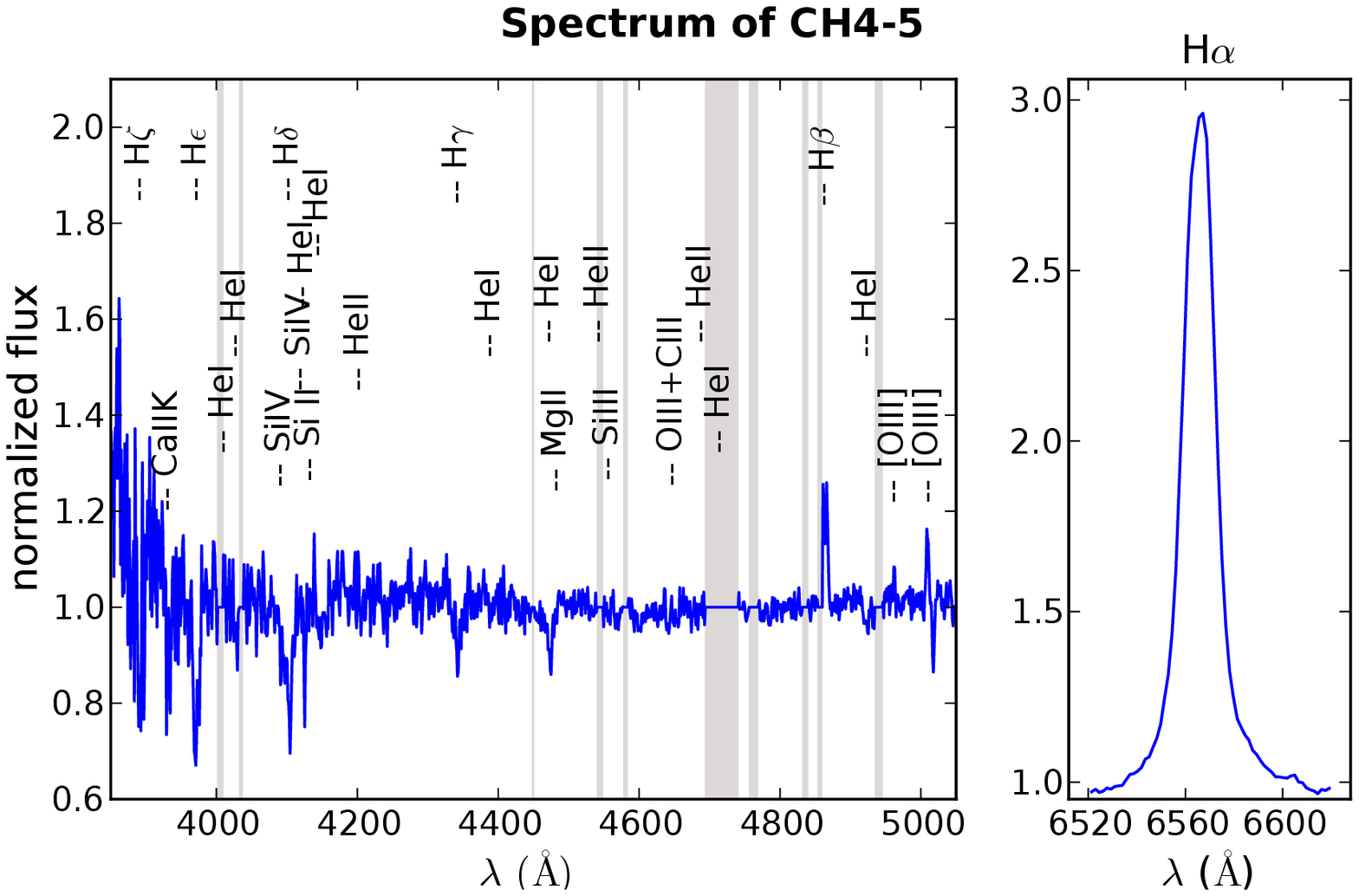} \\
\includegraphics[scale=0.59]{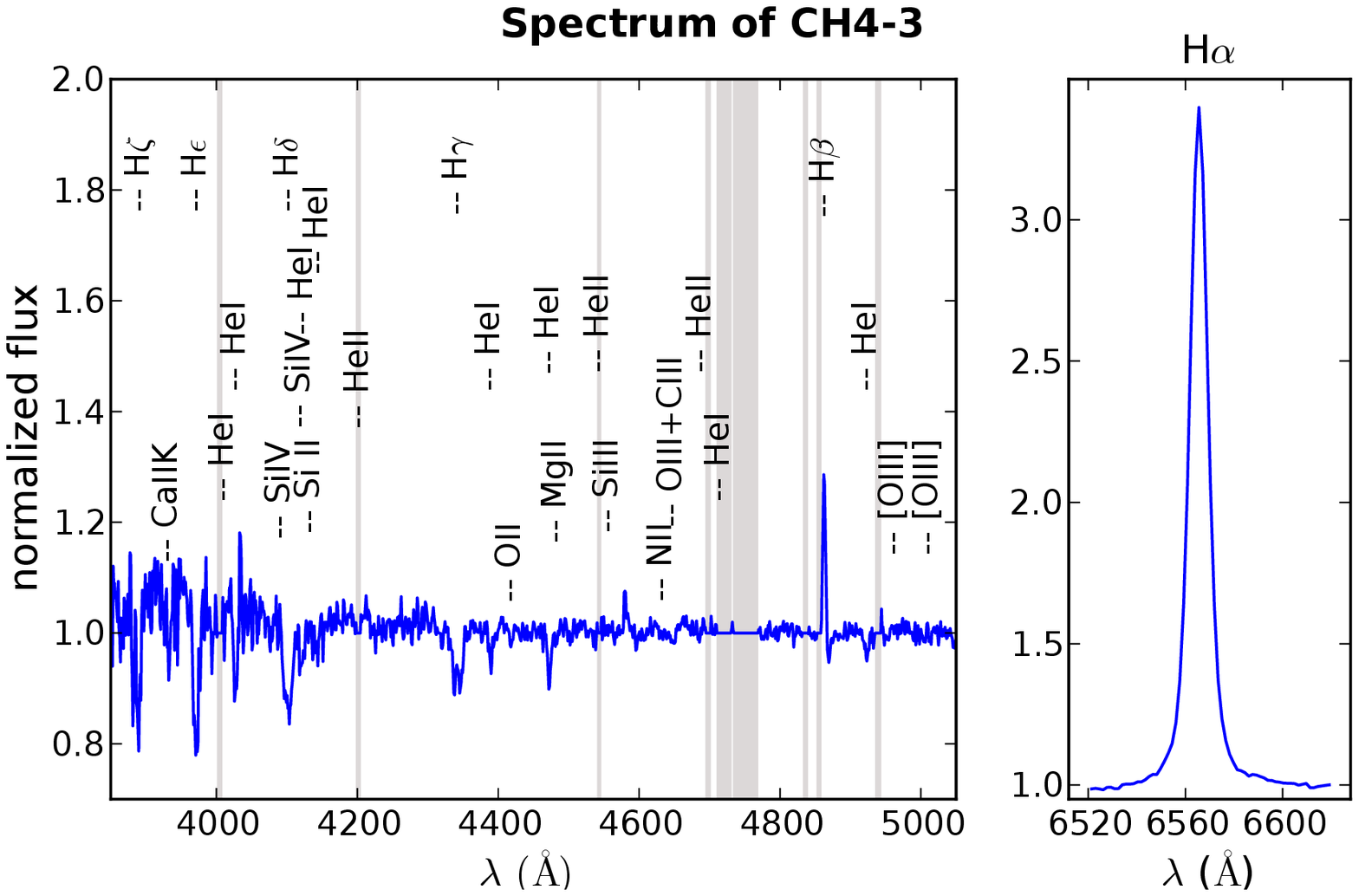} \\
\includegraphics[scale=0.59]{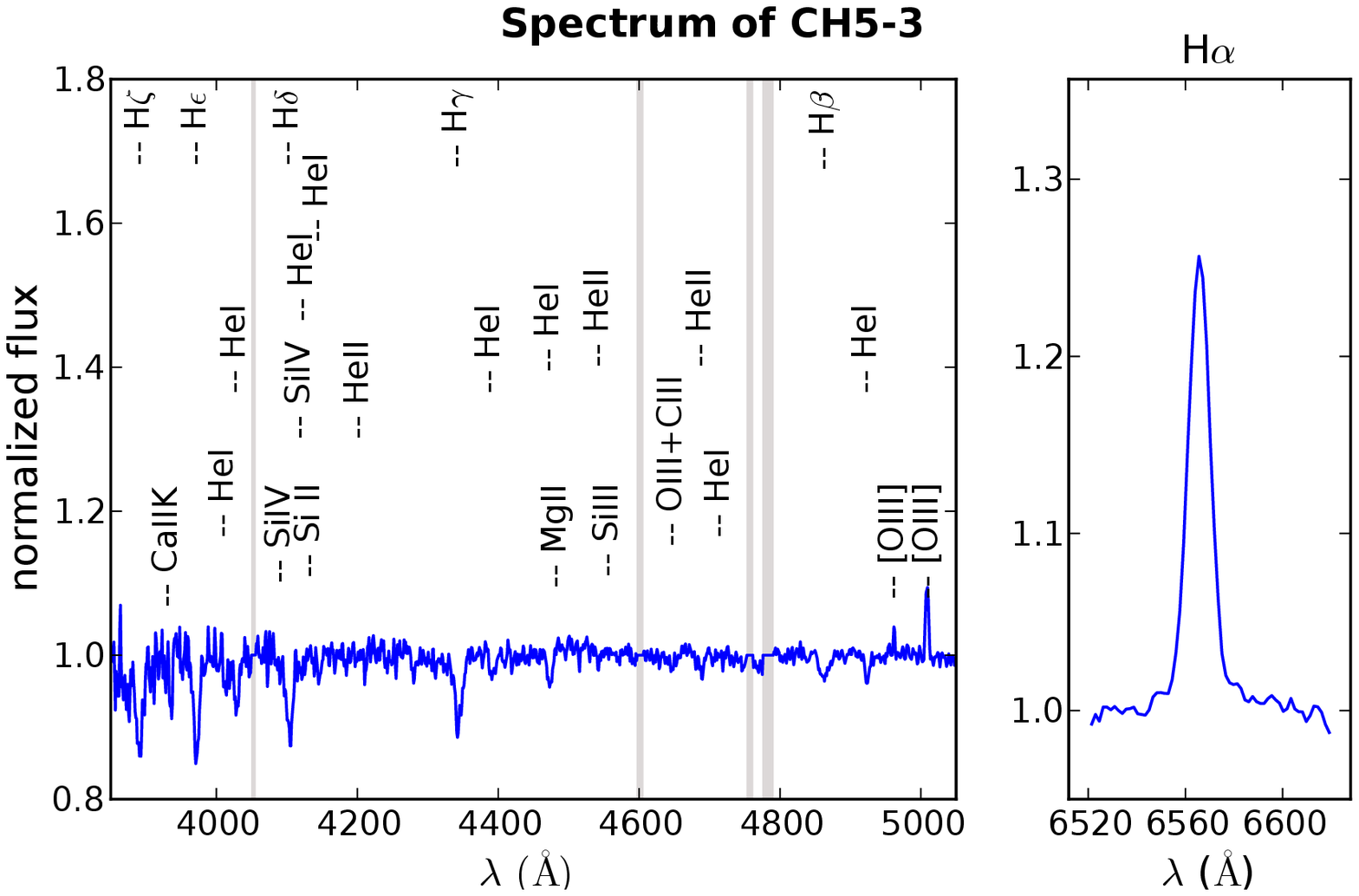} \\
\contcaption{}
\end{figure*}

\begin{figure*}
\includegraphics[scale=0.59]{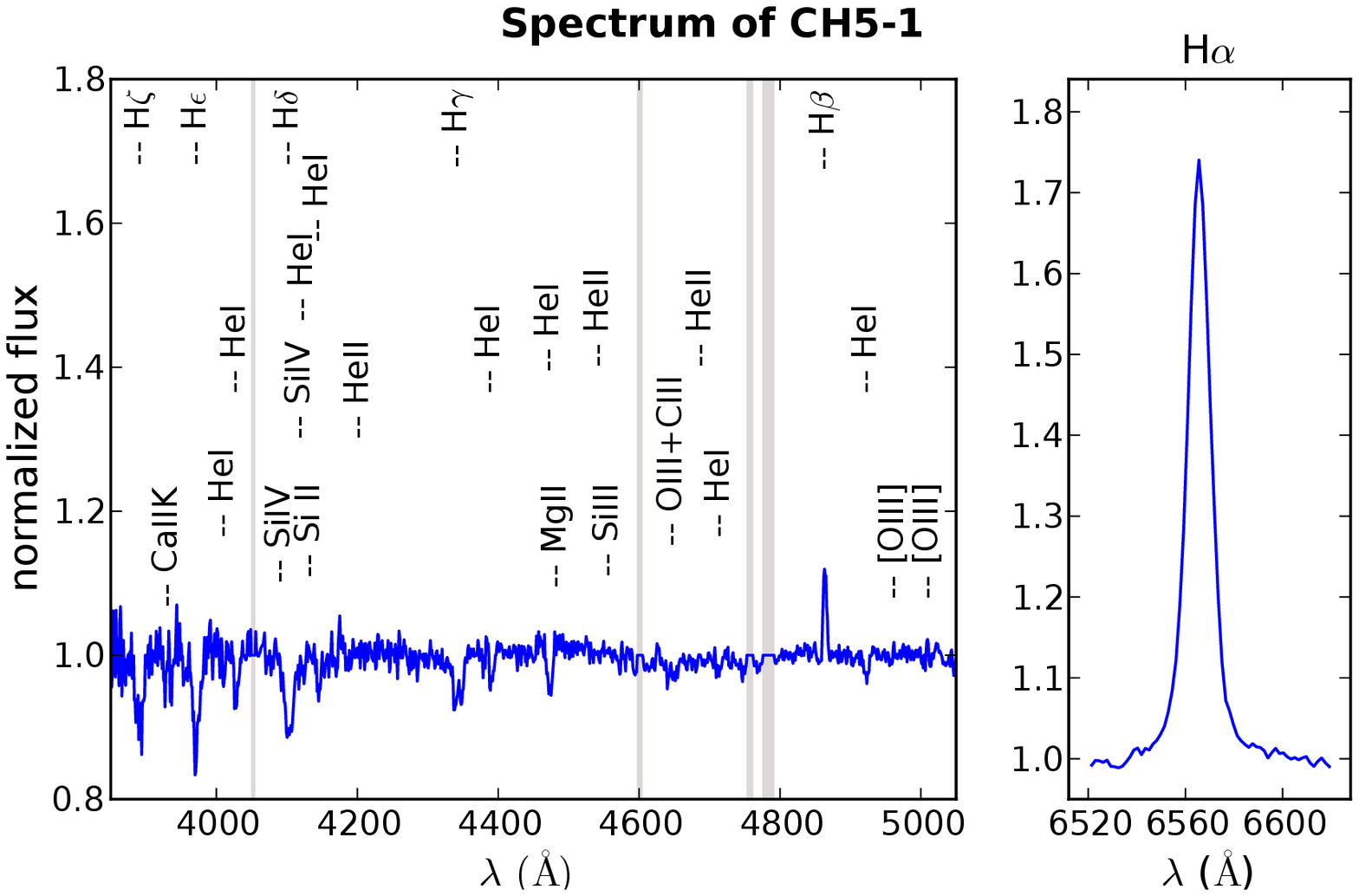} \\
\includegraphics[scale=0.59]{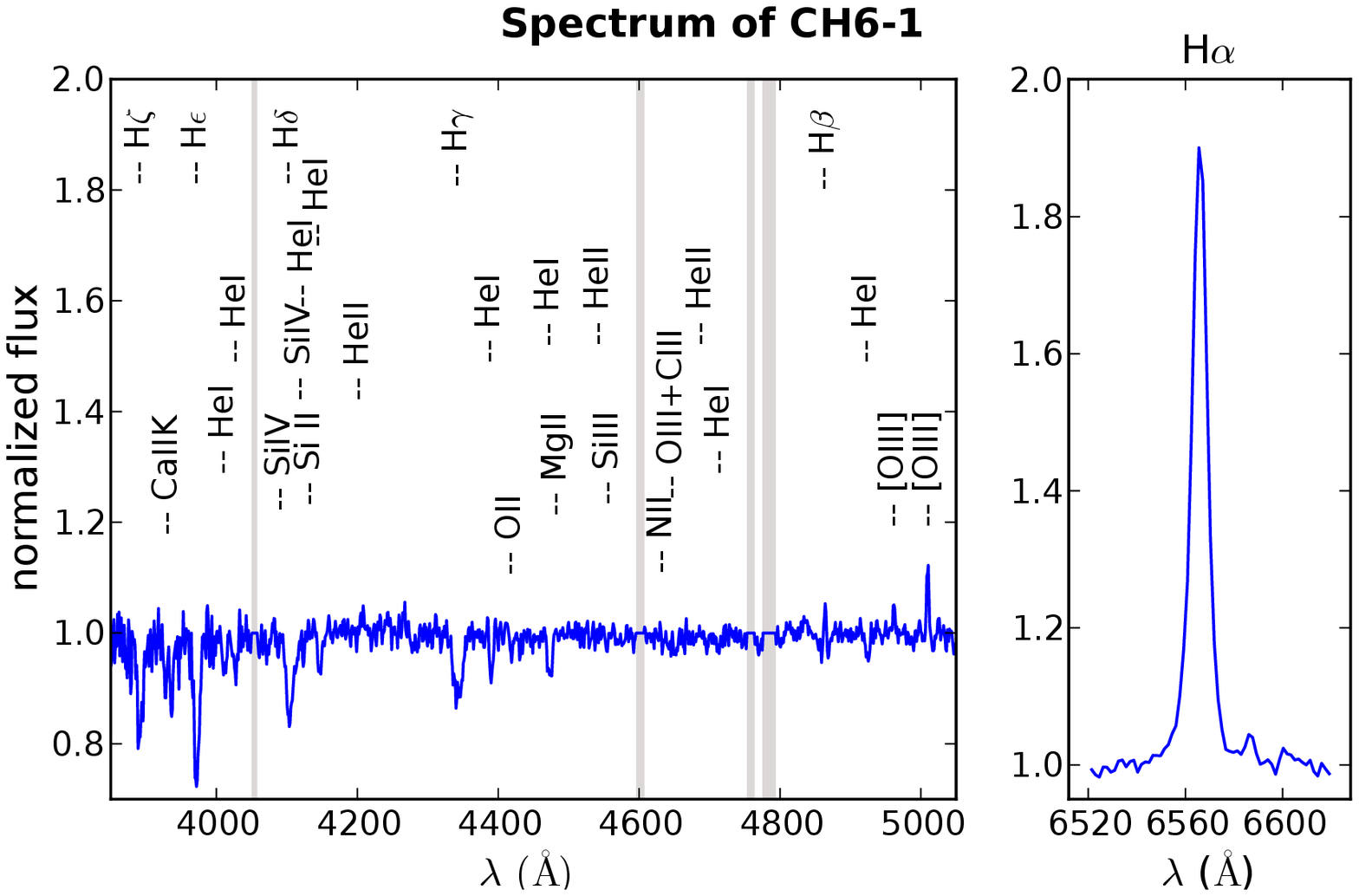} \\
\includegraphics[scale=0.59]{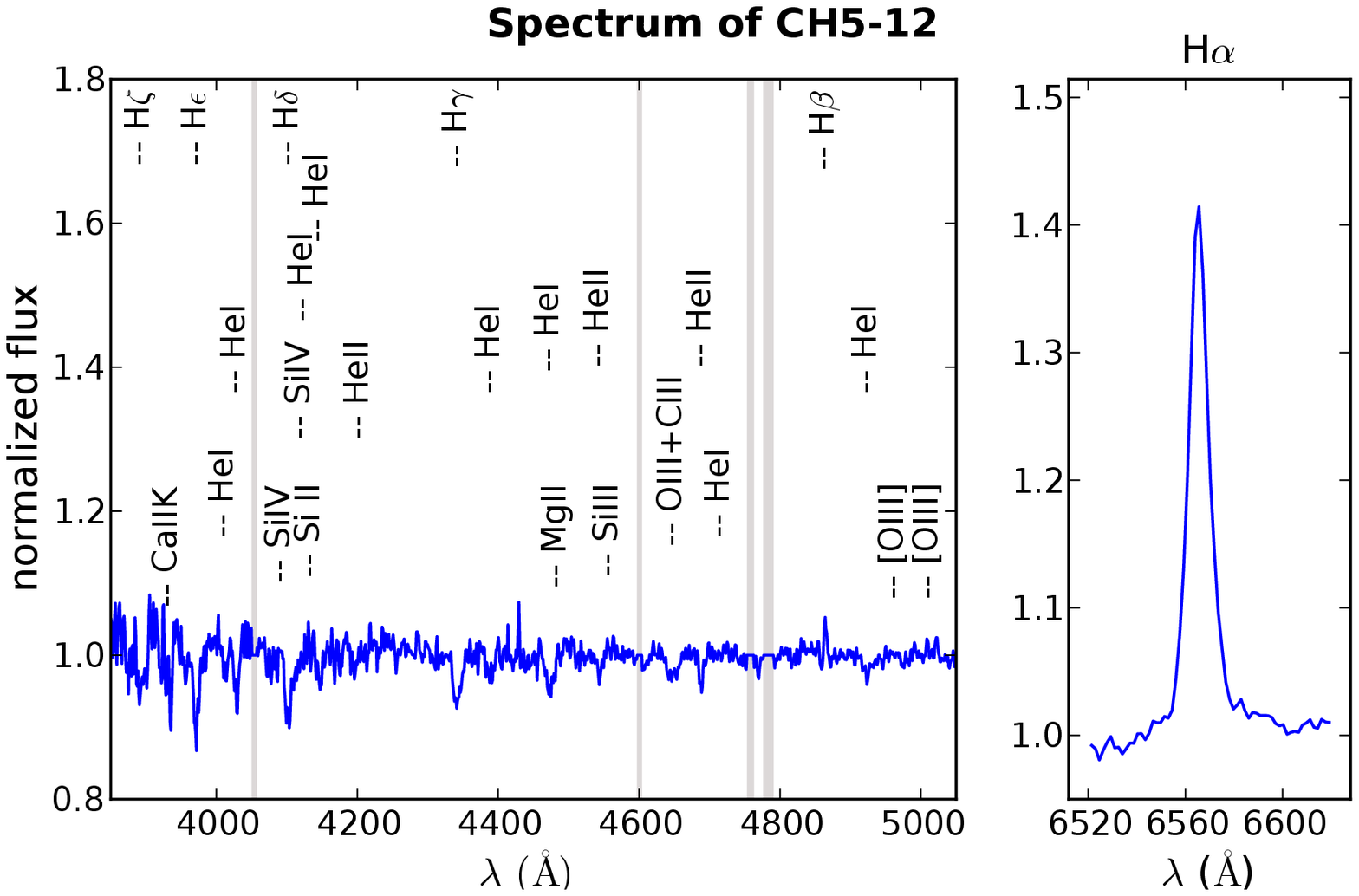} \\
\contcaption{}
\end{figure*}

\begin{figure*}
\includegraphics[scale=0.59]{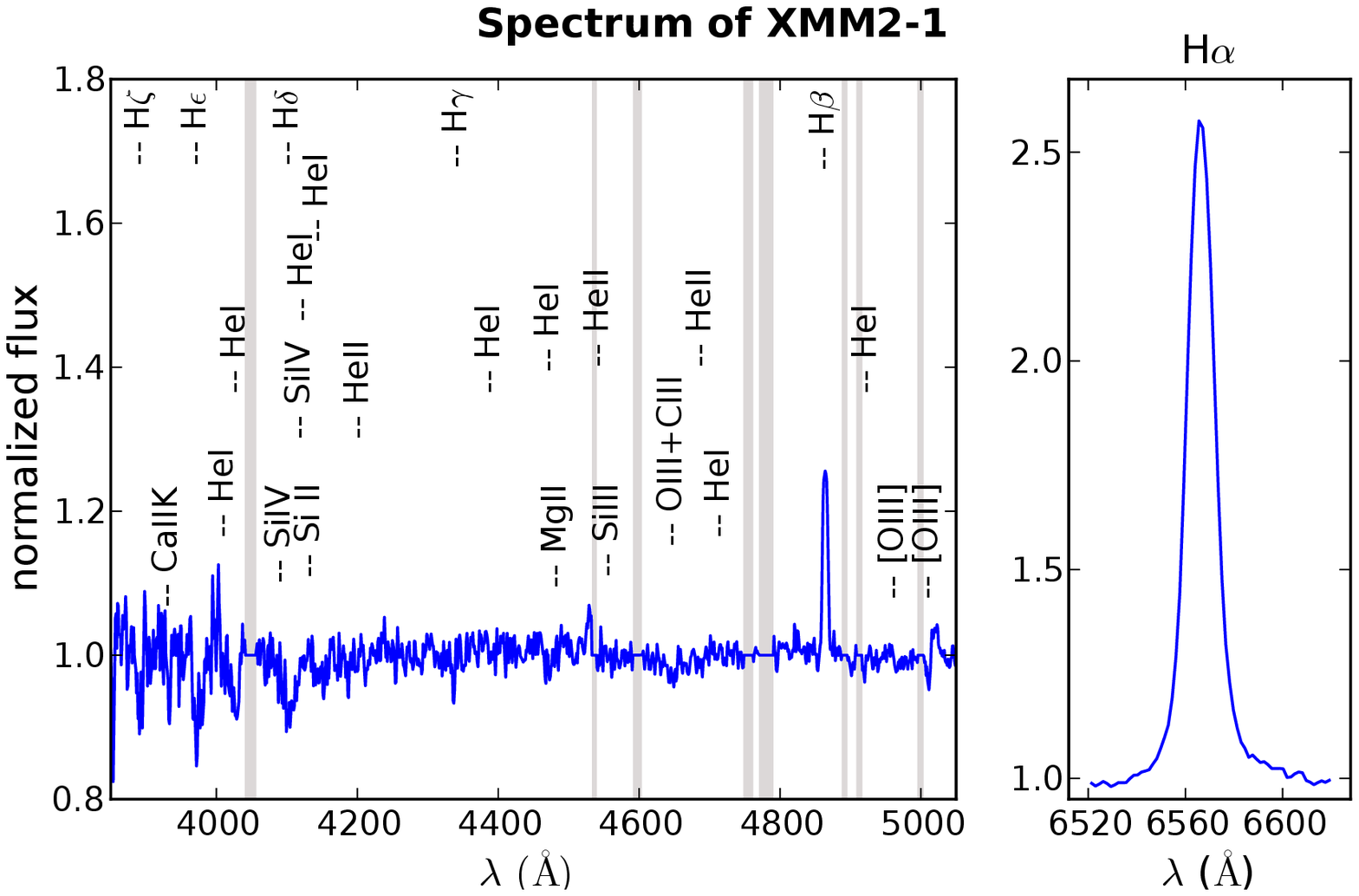} \\
\includegraphics[scale=0.59]{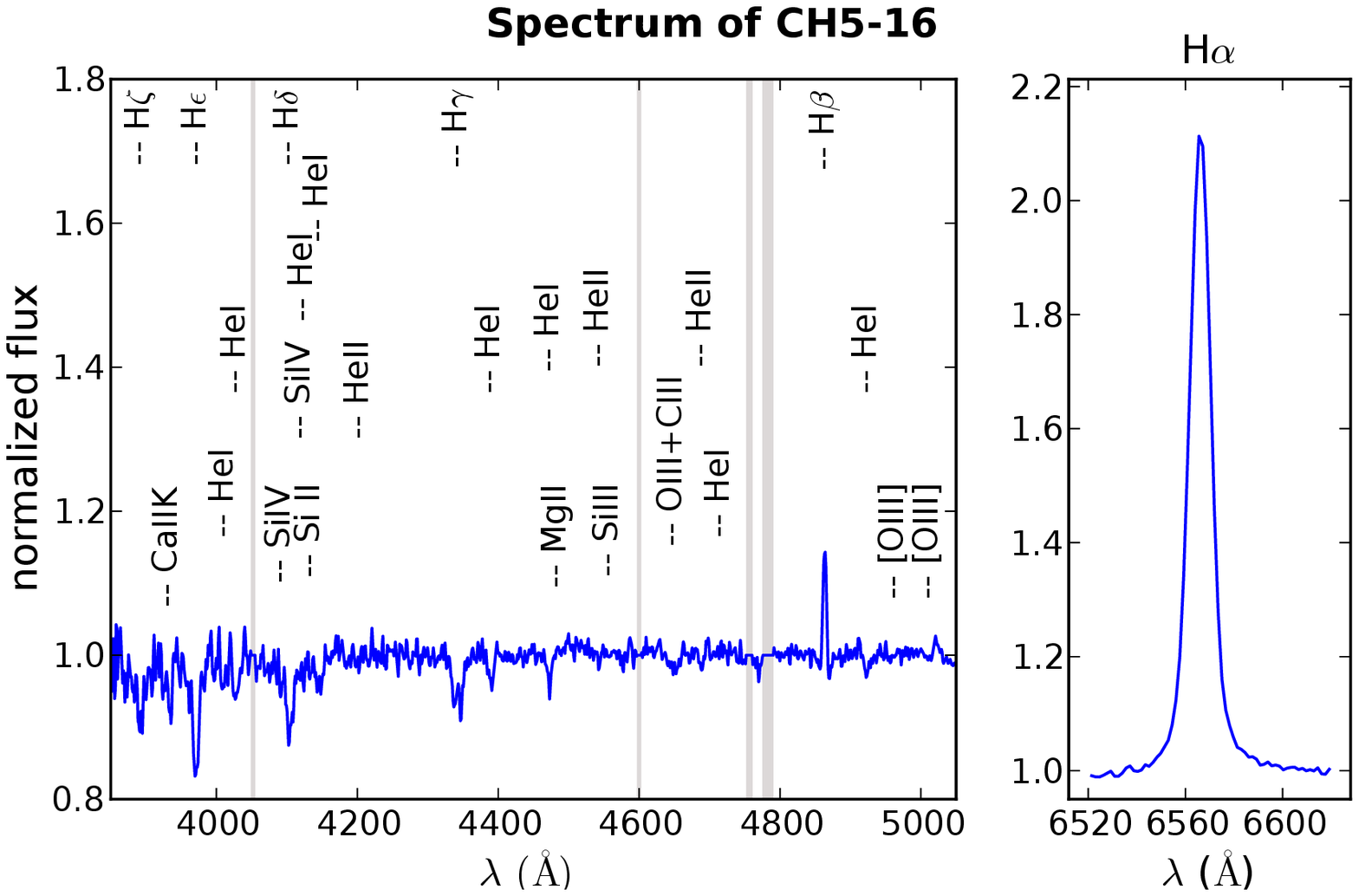} \\
\includegraphics[scale=0.59]{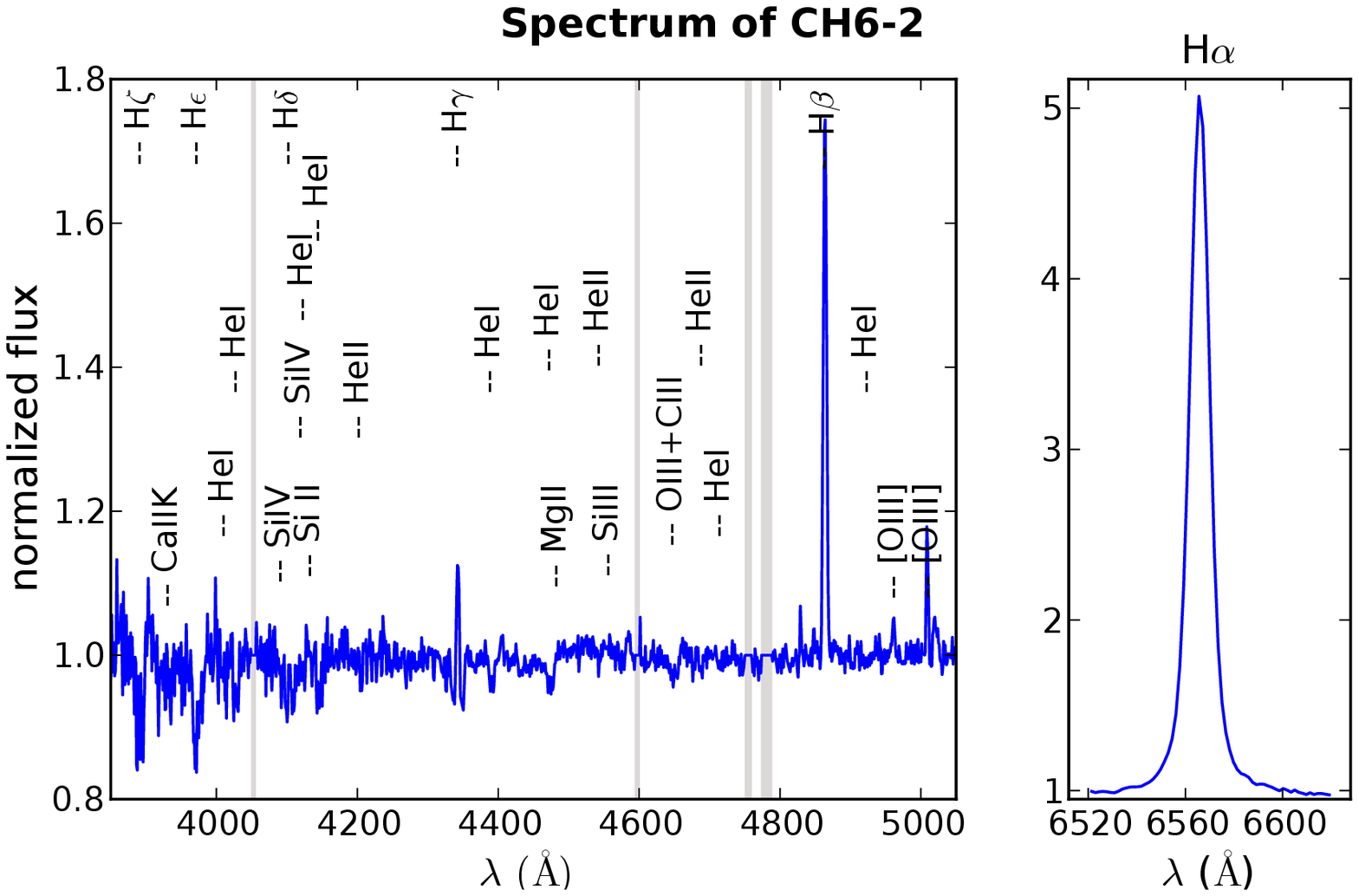} \\
\contcaption{}
\end{figure*}

\begin{figure*}
\includegraphics[scale=0.59]{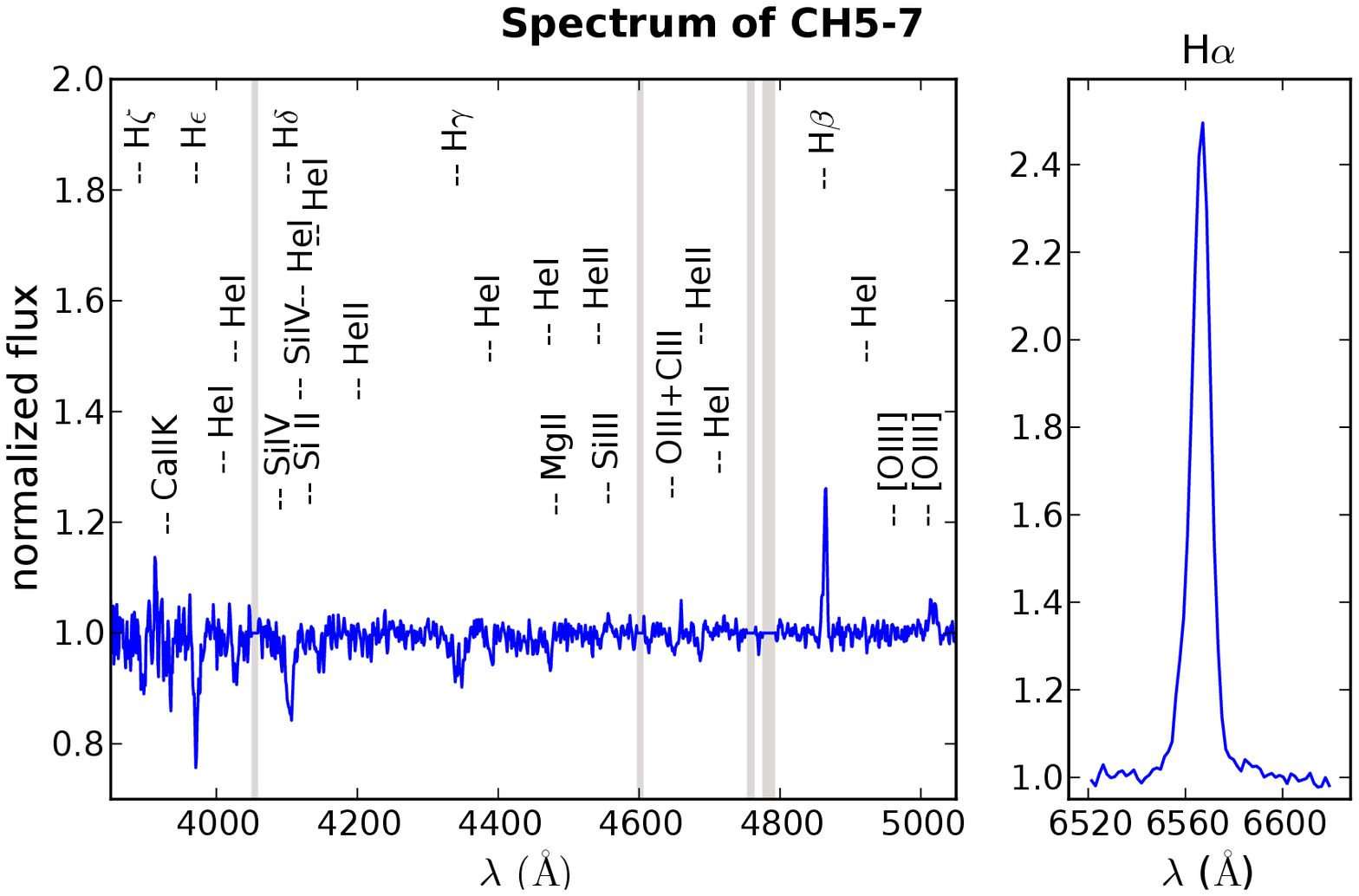} \\
\includegraphics[scale=0.59]{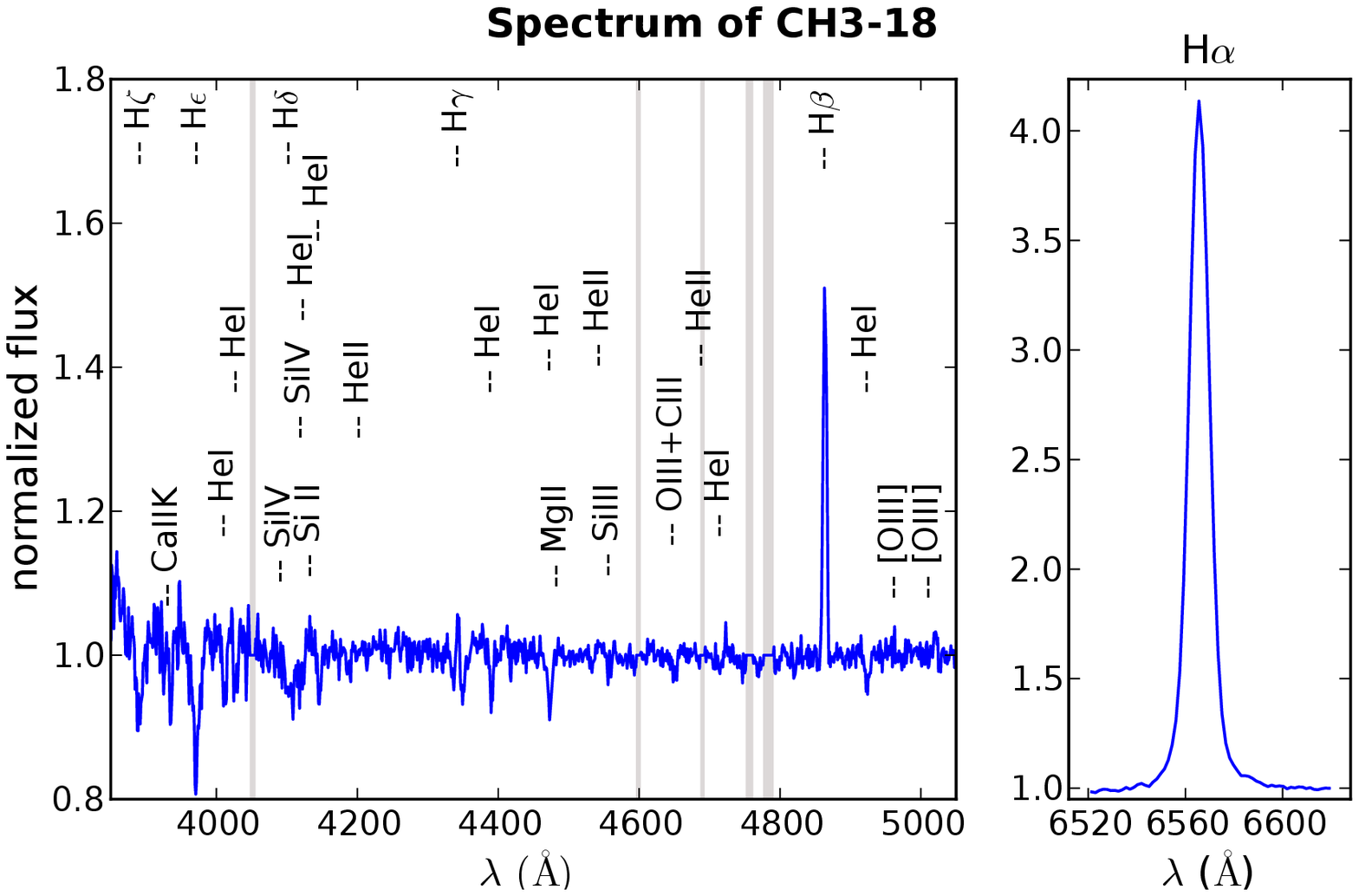} \\
\includegraphics[scale=0.59]{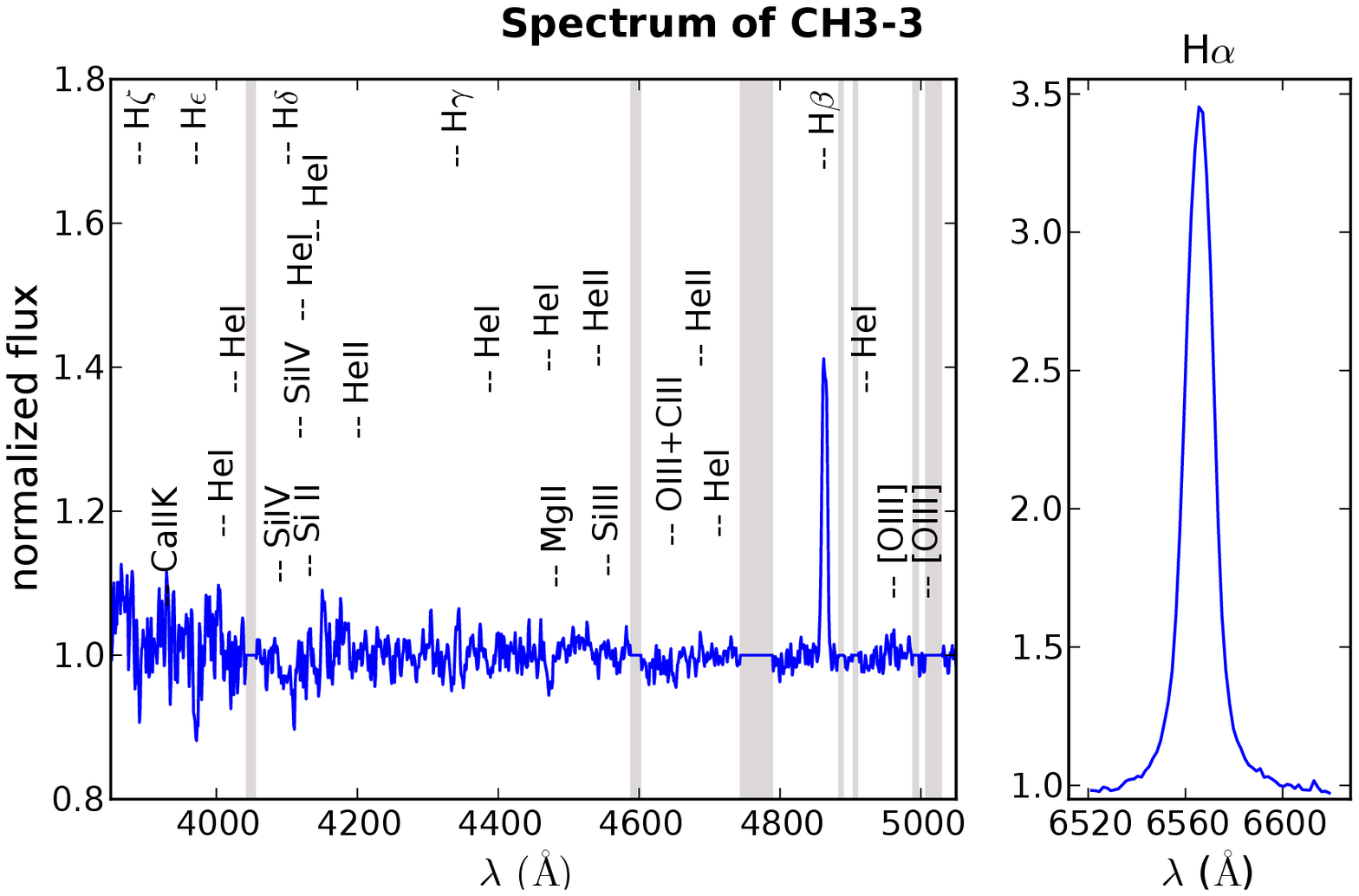} \\
\contcaption{}
\end{figure*}

\begin{itemize}

 \item \textit{CXOU J004814.15-731004.1 (source CH4-8) - classified as B1.5}

This source is classified as B1.5, in full agreement with \citet{Antoniou09}.

 \item \textit{CXOU J004903.37-725052.5 (source CH7-1) - classified as B1-B5}
 
The absence of both HeII $\lambda\lambda$4200, 4686 lines and of the MgII $\lambda$4481 line combined with the presence of the HeI $\lambda$4471 line, constrain the spectral type in the B1-B5 range, in agreement with  \citet{McBride08} who classified this source as $\sim$B3.

 \item \textit{CXOU J004913.57-731137.8 (source CH4-2) - classified as B3-B5}
 
For this source we obtained spectra on both nights (July 26 and September 19, 2008). The spectral range B3-B5  is determined by the absence of the HeII $\lambda\lambda$4200, 4541, and 4686 lines, and the OII+CIII $\lambda$4640-4650 blend, as well as the stronger HeI $\lambda$4471 line compared to the MgII $\lambda$4481 line. The resulting spectral type is later than the previous classification of B1.5 in \citet{Antoniou09}, where the OII+CIII $\lambda$4640-4650 blend and SiIV $\lambda\lambda$4088, 4116 lines were identified. 

 \item \textit{CXOU J004929.74-731058.5 (source CH4-5) - classified as B1-B5}
 
 The absence of the HeII $\lambda\lambda$4200, 4686 lines, and the combination with the clear presence of the HeI $\lambda$4471 line and the absence of MgII $\lambda$4481, indicates a spectral type in the B1-B5 range. \citet{Antoniou09} provided a spectral type of B1, based on the presence of the OII+CIII $\lambda$4640-4650 blend and SiIV $\lambda\lambda$ 4088, 4116 lines, which are not detected in our deeper, higher resolution spectra. 

 \item \textit{CXOU J005057.16-731007.9 (source CH4-3) - classified as B1-B3}
 
Source CH4-3 was also observed both nights (July 26 and September 19, 2008). The absence of the HeII  $\lambda\lambda$4200, 4686 lines and the presence of the OII $\lambda$4415-4417 and OII+CIII $\lambda$4640-4650 blend limit the spectral type in the B1-B3 range. This source has been classified previously as B0.5 by \citet{Antoniou09} due to the weak presence of the HeII  $\lambda$4686 line, which is not detected in our spectra.

 \item \textit{CXOU J005153.16-723148.8 (source CH5-3) - classified as B0.5}
 
Given the weak HeII $\lambda$4686 line, and the absence of the HeII $\lambda\lambda$4200, 4541 lines, all the criteria for a B0.5 star are fulfilled, resulting in a little later type than the previous classification of O9.5-B0 by \citet{McBride08}, although in the latter work there are no details about the specific lines that led to this classification.

 \item \textit{CXOU J005205.61-722604.4 (source CH5-1) - classified as B3-B5}
 
The combination of the presence and relative strength of the HeI $\lambda\lambda$4009, 4026, and 4144 lines, and the absence of the MgII $\lambda$4481 line limits the spectral type to the B3-B5 range. This is later than the previous classification of B1-B1.5 by \citet{McBride08}, for which no details about the specific lines are presented.

 \item \textit{CXOU J005208.95-723803.5 (source CH6-1) - classified as B1-B3}
 
 This source is classified as B1-B3, in full agreement with \citet{Antoniou09}.

 \item \textit{CXOU J005245.04-722843.6 (source CH5-12) - classified as B0}

The clear presence of the HeII $\lambda\lambda$4541, 4686 lines combined with the fact that the HeII $\lambda$4200 is absent constrains the spectral type to B0, since for earlier types this line is of comparable strength to the HeII $\lambda$4541 line. Thus, all criteria for a B0 star are fulfilled, which improves the previous wider classification of O9-B0 by \citet{McBride08}.

 \item \textit{XMMU J005255.1-715809 (source XMM2-1) - classified as B1-B3}

The absence of the HeII $\lambda\lambda$4200, 4541, 4686 lines and the presence of the OII+CIII $\lambda$4640-4650 blend, allow us to classify source XMM2-1 as B1-B3, which is in marginal agreement with the previous classification of B0-B1 by \citet{McBride08}.  

 \item \textit{CXOU J005355.25-722645.8 (source CH5-16) - classified as B0}
 
The HeII $\lambda\lambda$4541, 4686 lines are present, but not the HeII $\lambda$4200 line, indicating that this is a B0 star (earlier types display the HeII $\lambda$4200 line in at least comparable strength to the HeII $\lambda$4541 line). \citet{Antoniou09} have classified this source as B0.5 based on the absence of the HeII $\lambda$4541 line, in contrast to our deeper spectrum, where it is clearly seen. 

 \item \textit{CXOU J005455.78-724510.7 (source CH6-2) - classified as B1.5-B3}
 
The presence of the HeII $\lambda\lambda$4200, 4686 lines and the absence of SiIV $\lambda$4116 limit the spectral range to an as early type as B1.5, while the presence of the OII+CIII $\lambda$4640-4650 blend limits the spectral type to no later than B3. This is marginally consistent with the previous classification of B1-B1.5 by \citet{Antoniou09}, based on the presence of the SiIV $\lambda\lambda$ 4088, 4116 lines, which however are not detected in our deeper spectra.    

 \item \textit{CXOU J005456.34-722648.4 (source CH5-7) - classified as B0.5}
 
The absence of the HeII $\lambda$4200 line and the presence of the HeII $\lambda$4686 line suggest a spectral type of B0.5. The slightly earlier classification of B0 by \citet{Antoniou09} was based on the presence of the HeII $\lambda$4200 line, which is not seen in our deeper spectrum. 

 \item \textit{CXOU J005605.42-722159.3 (source CH3-18) - classified as B2}
 
The SiIII $\lambda$4553 line is clearly present but without any sign of the SiIV $\lambda\lambda$4088, 4116 lines. This combination is in agreement with the criteria set for B2 stars. This classification is later than the previous classification of B1 by \citet{McBride08}.  

 \item \textit{CXOU J005736.00-721933.9 (source CH3-3) - classified as B1-B5}
 
The absence of the HeII $\lambda\lambda$4200, 4686 lines, in combination with a clearly present HeI $\lambda$4471 line and the absence of the MgII $\lambda$4481 line, suggest a spectral range of B1-B5. This classification result is in agreement with the also broad classification of B0-B4 by \citet{Antoniou09}.     

\end{itemize}

This paper has been typeset from a \TeX/\LaTeX\ file prepared by the author.

\end{document}